\documentclass[a4paper,11pt]{article}
\pdfoutput=1 

\usepackage{jinstpub} 

\usepackage[inter-unit-product=\ensuremath{{}\cdot{}}]{siunitx}
\usepackage{graphicx,subcaption}
\usepackage{amsmath}
\usepackage{booktabs}
 \usepackage{multirow}
\usepackage{hyperref}
\hypersetup{
    colorlinks=true,
    linkcolor=blue,
    filecolor=magenta,      
    urlcolor=cyan,
}
 
\title{\boldmath A Prototype Scintillating Fibre Beam Profile Monitor for Ion Therapy Beams}

\author[a,1]{B.D. Leverington,\note{Corresponding author.}}
\author[a]{M. Dziewiecki,}
\author[a]{L. Renner,}
\author[a,2]{R. Runze,\note{Currently attending Northwestern University.}}

\affiliation[a]{Physikalisches Institut, Universit\"at D-69120 Heidelberg, Germany}

\emailAdd{leverington@physi.uni-heidelberg.de}

\abstract{A prototype plastic scintillating fibre based beam profile monitor was tested at the Heidelberg Ion Therapy Centre / \textit{Heidelberg Ionenstrahl Therapiezentrum} (HIT)  in 2016 to determine its beam property reconstruction performance and the feasibility of further developing an expanded system. 
At HIT protons, helium, carbon, and oxygen ions are available for therapy and experiments.  The beam can be scanned in two dimensions using fast deflection magnets. A tracking system is used to monitor beam position and to adjust scanning magnet currents online.
 A new detector system with a finer granularity and without the drift time delay of the current MWPC system with a similar amount of material along the beamline would prove valuable in patient treatment. 
 The sensitive detector components in the tested prototype detector are double-clad Kuraray SCSF-78MJ scintillating fibres with a diameter of 0.250~mm wound as a thin multi-layer ribbon. The scintillation light is detected at the end of the ribbon with Hamamatsu S11865-64  photodiode arrays with a pitch of 0.8~mm. Commercial or readily available readout electronics have been used to evaluate the system feasibility. 

 The results shown in this paper include the linearity with respect to beam intensity, the RMS of the beam intensity as measured by two planes, along with the RMS of the mean position, and the measured beam width RMS. The Signal-to-Noise ratio of the current system is also measured as an indicator of potential performance. Additionally, the non-linear light yield of the scintillating fibres as measured by the photodiode arrays is compared to two models which describe the light yield as a function of the ion stopping power and Lorentz $\beta$.}


\begin{document}
\maketitle
\flushbottom

\section{Introduction}

\begin{figure}[htbp]
\centering
  \includegraphics[width=0.5\linewidth]{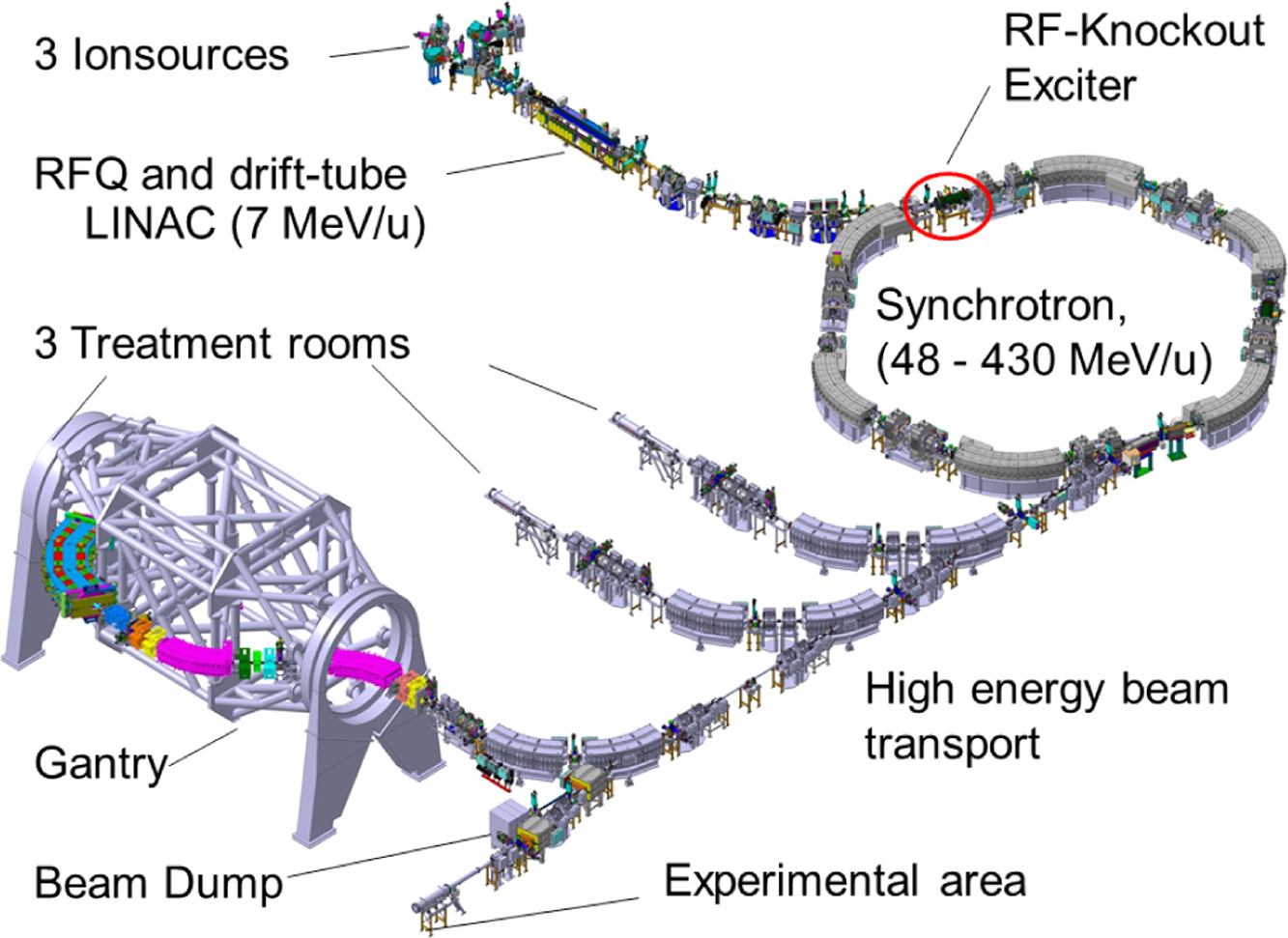}
  \caption{ The HIT accelerator synchrotron complex with patient treatment rooms and an experimental area. Figure from \cite{bib:Ondreka}.}
  \label{fig:hitaccel}
\end{figure}

The Heidelberg Ionenstrahl Therapiezentrum (HIT) is a radiation therapy facility
located on the campus of the Heidelberg University Hospital. It is one of the first synchrotron facilities in Europe
providing proton as well as heavy ion beams and is designed to support a raster scanning dose delivery method \cite{bib:Haberer, bib:Haberer2}. An overview of the accelerator system at HIT is shown in Figure~\ref{fig:hitaccel}. Protons, helium, carbon and oxygen ions are available from  three ion sources\footnote{Each  source can supply every ion type. Multiple sources allows for rapid switching without delays due to restarting the source.} and are accelerated to 7~MeV/u in the Injector Linac. The final accelerator stage takes place in a 65~m circumference synchrotron. As soon as the particles have reached the desired energy they are guided by the high energy beam transport line to one of the three patient treatment rooms or an experimental area. The ranges of energies and intensities available for each ion type are shown in Table~\ref{tab:HITbeam}. 

\begin{table}[htbp]
\centering
\begin{tabular}{l|c|c|c|c|}
                     & Protons                       & Helium                      & Carbon                        & Oxygen                        \\ \hline
Energy [MeV/u]       & \numrange{48}{221}                    & \numrange{51}{221}                  & \numrange{89}{430}                    & \numrange{104}{515}                   \\
Intensity [s$^{-1}$] & \num{8e7}--\num{3.2e9} & \num{2e7}--\num{8e8} & \num{2e6}--\num{8e7} & \num{1e6}--\num{4e7} \\ \hline
\end{tabular}
\caption{The energies and intensities available at the HIT Clinic. The energy range is divided into 255 possible settings, E1-E255. There are typically 10 different intensity settings available for use, I1-I10.}
\label{tab:HITbeam}
\end{table}

The beam in the treatment and experimental areas can be steered over a \SI{20x20}{\cm\square} area such that a pre-determined integrated fluence of particles can be deposited over a given area. Patient treatment plans are typically divided into voxels on the order of \si{\cubic\milli\metre} with the penetration depth of the beam determined by the location of the Bragg peak.  The beam position, width and intensity are monitored online by ionisation and multi-wire proportional chambers (IC and MWPC) placed 1-1.5~m upstream from the treatment iso-centre. A treatment plan is typically prepared such that the desired ionising radiation dose is deposited in the tumor with minimal dose being applied outside the treatment region. The stability of the intensity and position of the beam are monitored  to ensure the correct dose is applied in each voxel, while also allowing for the intensity to be adjusted relative to the dose needed in each voxel \cite{bib:Schoemers}. 

In addition to patient treatment and clinical studies at the clinic, there is an ongoing effort to improve the accelerator facility as well. Currently, the maximimum intensity that can be applied at each raster point is limited by the time required to measure the position with the MWPC, approx. \SI{1}{\milli\sec}, with a deadtime of approximately \SI{150}{\micro\sec}\cite{bib:Schoemers}. A decrease in the total treatment time due to higher beam rates would have multiple impacts. In addition to reduced patient treatment times and the corresponding increase in the number of patients per day that could be treated, the uncertainties in the position of the target dose volume would also decrease; Internal motions of the patient (such as respiration) can lead to uncertainties in the locations of the applied dose. There is interest at HIT for a new tracking detector system which will allow for these improvements at the end of the lifetime of the current system. The desired performance specifications for such a detector are summarised in Table~\ref{tab:specs} and are required over the beam settings specified in Table~\ref{tab:HITbeam}.

\begin{table}[]
\centering
\caption{The desired specifications for a beam profile monitor at HIT.}
\label{tab:specs}
\begin{tabular}{@{}lll@{}}
\toprule
Requirement             & Value                   &   \\ \midrule
Beam Spot Size (FWHM)        & 1 -- 33 mm              &   \\ 
Beam Position Resolution     & \textless 0.2 mm                   &  \\
Beam Width Resolution        & \textless 0.4 mm                  &   \\
Readout Rate            & 4 -- 8 kHz              &   \\
Dead Time               & \textless \SI{250}{\mics} &   \\ 
Material in Active Area &   \textless 0.35 mm H$_2$O eq. / plane   &   \\ \bottomrule
\end{tabular}
\end{table}

\section{Detector}
The proof-of-concept beam profile monitor (BPM) described in this paper is part of the research effort to improve the facility. In addition to the delays from the MWPC readout and drift time, the granularity of the MWPC is limited by the \SI{2}{\milli\metre} wire spacing.    A system based on scintillating fibres for beam position monitoring offers some benefits over the present system. Aside from the relative simplicity of constructing the sensitive detector plane, its robustness, and that gas and high voltage are not required, the fibres offer intrinsic benefits. The scintillation decay time is typically on the order of a few nanoseconds; the spatial granularity is proportional to the fibre diameter, which can be as low as \SI{0.25}{\milli\metre} for plastic fibres; and the signal amplitude is proportional to the deposited energy in the fibre, though extensive research has shown this to be non-linear with increasing stopping power, dE/dx, for plastic scintillator, and will be examined in detail below. Photodiode arrays with a channel pitch of \SI{0.8}{\milli\metre} are used to measure the light output with integration times of the photodiode array down to \SI{100}{\mics}. Covering the desired acceptance of  \SI{20x20}{\cm\square} is also straight forward with scintillating fibres thanks to the production methods developed for the fibre tracker for the upgraded LHCb experiment at CERN \cite{bib:LHCbSciFiTDR}.  \\

The obvious drawback to this detector system is the additional material in the beam compared to the MWPC due to the minimal thickness achievable with a plastic fibre. The total material in the beamline at present is equivalent to \SI{2.9}{\mm} of water \cite{Parodi_2010}. The vacuum window, three ionisation chambers, and two MWPCs each contribute \SI{0.360}{\mm}, \SI{0.230}, and \SI{0.160}{\mm} of water equivalent (WE) material, respectively \cite{Parodi_2012}. Approximately \SI{1.1}{m} of air between the vacuum window and the isocenter contributes the additional \SI{1.53}{\mm}. An ideally thin station with two orthogonal scintillating fibre ribbons built from two staggered layers of \SI{0.250}{\mm} scintillating fibres without additional epoxy would have a combined thickness of \SI{0.6}{\mm} on average due to the staggering of the fibre layers. Replacing the MWPC with fibres would result in \SI{0.9}{\mm} or 30\% extra WE material, though lacking the argon, tungsten or aluminium of the other chambers. Two-layer scintillating fibre stations \textit{with epoxy} would contribute almost 60\% more WE material and is likely unacceptable. A preliminary study using GEANT4 \cite{Agostinelli:2002hh} to estimate the effect on the lateral spread of the beam is presented in Section~\ref{section:geant}. \\

  The measurements presented in this article aim to determine the initial feasibility of the scintillating fibre-based detector over the broad range of intensities and energies of the HIT ion beam by examining the dynamic range of the system along with the beam reconstruction precision. If the results are within reach of the proposed specifications, the authors will proceed with further development in order to surpass the proposed specifications. To this end, this feasibility study was performed using available components, and while not yet optimised in terms of readout electronics, the preliminary system should provide enough information to determine the intrinsic performance characteristics.


\subsection{Scintillating fibres}\label{sec:fibre}
The scintillating fibre ribbon used in this research were produced originally by the LCHb Scintillating Fibre Tracker collaboration. The fibres are Kuraray SCSF-78MJ  \SI{0.25}{\milli\metre} diameter double-clad fibres wound into a multi-layer ribbon on a custom machine~\cite{bib:SciFiEDR}. The emission spectra of these fibres\footnote{While not directly confirmed by the manufacturer, the polystyrene core is known to be doped with p-terphynyl (1\% w/w) as the primary dopant and TPB (tetraphynyl-butadiene) (0.1\% w/w) as the wavelength shifter. The cladding is PMMA based. } peaks at 450~nm .  The fibres have been bonded with a titanium dioxide(20\% w/w) loaded two-component epoxy\footnote{Epotek 301-2}. The overall thickness of the five layer mat is approximately \SI{1.2}{\milli\metre}, larger than the \SI{0.8}{\milli\metre} acceptance of the photodiode channel. This is also significantly more material than specified in Table~\ref{tab:specs}. If the dynamic range and sensitivity of the detector is feasible, 2-layers mats without glue will be developed, reducing the material by a factor of five (discussed further in detail in Sec.\ref{section:geant}).  However, the thicker mats used in this measurement will still provide the information needed to evaluate the dynamic range and other performance characteristics. \\

\begin{figure}[htbp]
\centering
  \includegraphics[trim={0 0cm 0 0cm},clip, width=0.5\linewidth]{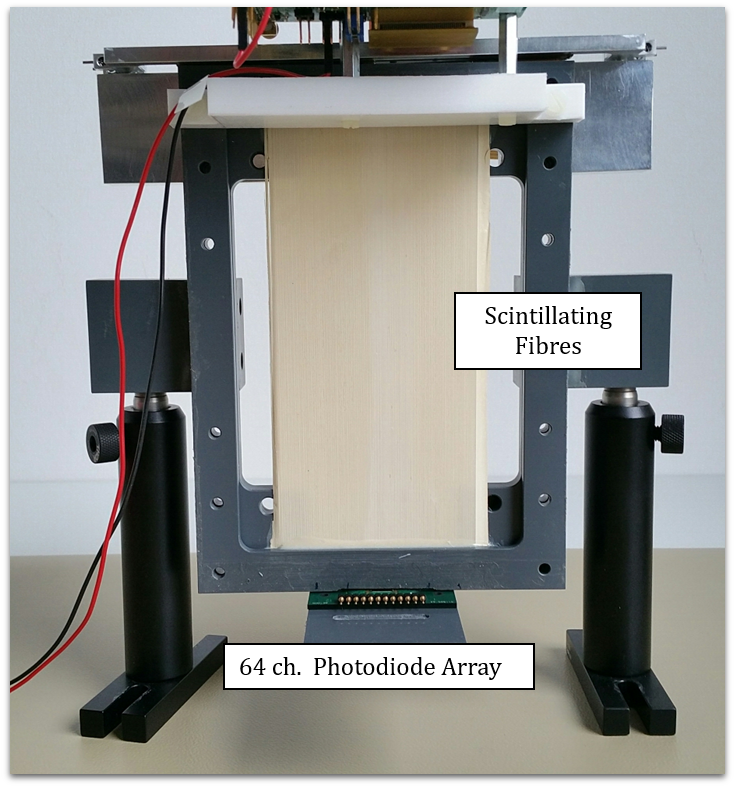}
  \caption{ A photo of the simple proof-of-concept detector with a single  \SI{7}{\centi\metre} wide mat with a single photodiode array attached at the bottom. The readout electronics have been removed.  The fibres are aligned in the vertical direction. }
  \label{fig:planeschem}
\end{figure}

   The ribbons tested were \SI{20}{\centi\metre} long\footnote{The maximum width of the mat is currently limited to  \SI{13}{\centi\metre} by the production wheel used to wind the mat, and the length can be as long as \SI{2.5}{\metre}.} and the width covered by the photosensor was \SI{51.2}{\mm}. The standard horizontal fibre pitch (\SI{0.275}{\milli\metre}) is larger than the fibre diameter to accommodate diameter variations. The proof-of-concept detectors studied here were produced by cutting a standard \SI{2.5}{\metre} LHCb fibre ribbon that was available to us at the time in order to produce the shorter detector planes. Each ribbon is bonded between two PVC frames and the ends are milled with a single-point diamond blade to produce a high quality optical finish. The photodetectors are mounted at the end of the PVC frame for optical readout of the fibre mat.  Mirrors have not been used here but are considered for use in the future. A photo of the tested prototype is shown in Figure~\ref{fig:planeschem}.


\subsection{Photodiode arrays}\label{sec:photodiode}

The beam width at the current tracking detector (MWPC) locations can be as small as \SI{1}{\milli\metre} FWHM such that the granularity of the tracking planes should be the same or smaller. The Hamamatsu S11865-64 photodiode arrays used here have 64 channels with a pitch of \SI{0.8}{\milli\metre}, covering \SI{51.2}{\milli\metre} per array\footnote{A 128-ch version of the photodiode array is available, but the intent will be to readout multiple arrays in the future in parallel. For a similar clock speed, the readout rate per array is two times faster for 64-ch arrays.}.  The pixel sizes are \SI{0.7x0.8}{\mm\square}~(WxH). 
The photosensitivity of the array peaks at \SI{700}{\nm} and matches well with the wavelength emission of the scintillating fibre. 

  The analog signal processing is performed by an onboard CMOS chip. The signal from the channel is integrated simultaneously and readout sequentially. The maximum readout rate is \SI{12.5}{\kilo\Hz} per 64~channel array. The C9118 driver circuit from Hamamatsu has been used to interface the array to the data acquisition system. The  video output, which corresponds to the integrated charge of each channel, has been measured using the Heidelberg LogicBox system with a single channel SU735-2 12-bit ADC module \cite{bib:logicbox}. The integration windows can be adjusted by the timing of control signals, and values from \SI{25}{\mics} up to \SI{2500}{\mics} are possible.  Additionally, a gain can be selected from two values, low or high, which differ by a factor of two.  The data transfer from the LogicBox to a PC uses the USB 2.0 protocol, which creates a readout rate limitation of 1-3 kHz.  The future improvements to this system will attempt to achieve a readout rate for the complete system greater than \SI{10}{\kilo\Hz} with a dead time of \SI{5}{\micro\sec} for multiple arrays. This is a significant improvement over the gas detectors with dead times (ion drift times) of more than \SI{100}{\micro\sec}.

It should be noted that  a similar prototype detector to the one constructed here with the same photodiodes\footnote{The 128-ch version of the photodiode was used here.} and similar readout electronics has been developed previously by Rojatti et al. \cite{Rojatti:2015npp} for the CNAO proton and carbon ion beamline in Italy but with \SI{0.5}{\milli\metre} square fibres, covering  a much smaller area (\SI{6x6}{\cm\square}),  a slower readout rate of \SI{300}{\Hz}, and a much longer integration time of \SI{2.5}{\milli\sec}.  Comparable performance results from the CNAO detector have not been published.

 \section{Results}

It was not clear from the outset how the detector would respond over the range of energies and intensities available at HIT for the different ion types, as the experience to date with these fibre ribbons was based on single minimum ionising -like particles and photon counting photodetectors. The results presented here are measurements scanning over intensity and energy, to observe the response of the detector system test for linearity and dynamic range of the photodiodes coupled to the scintillating fibres and with the LogicBox readout. \\

A lab measurement was first made with a UV source with subsequent measurements performed in the ion beamline. In the beamline, the detector has been placed at the treatment iso-center, approximately \SI{1}{m} from the exit window of the nozzle, indicated by the crossing points of the lasers in the HIT experimental room. For all beam conditions, the smallest available focus setting, F1, was chosen. The absolute FWHM of the lateral beam spread at the treatment isocentre in air depends on the beam energy.  Data was collected at a rate of \SI{3}{\kilo\hertz} during 2--3 full spills from the synchrotron which have a length of approximately five seconds each. The structure of the spills is described in detail in Ref.~\cite{bib:Schoemers}.

\subsection{Linearity with a UV light source}
\begin{figure}[htbp]
\centering
  \includegraphics[width=0.5\linewidth]{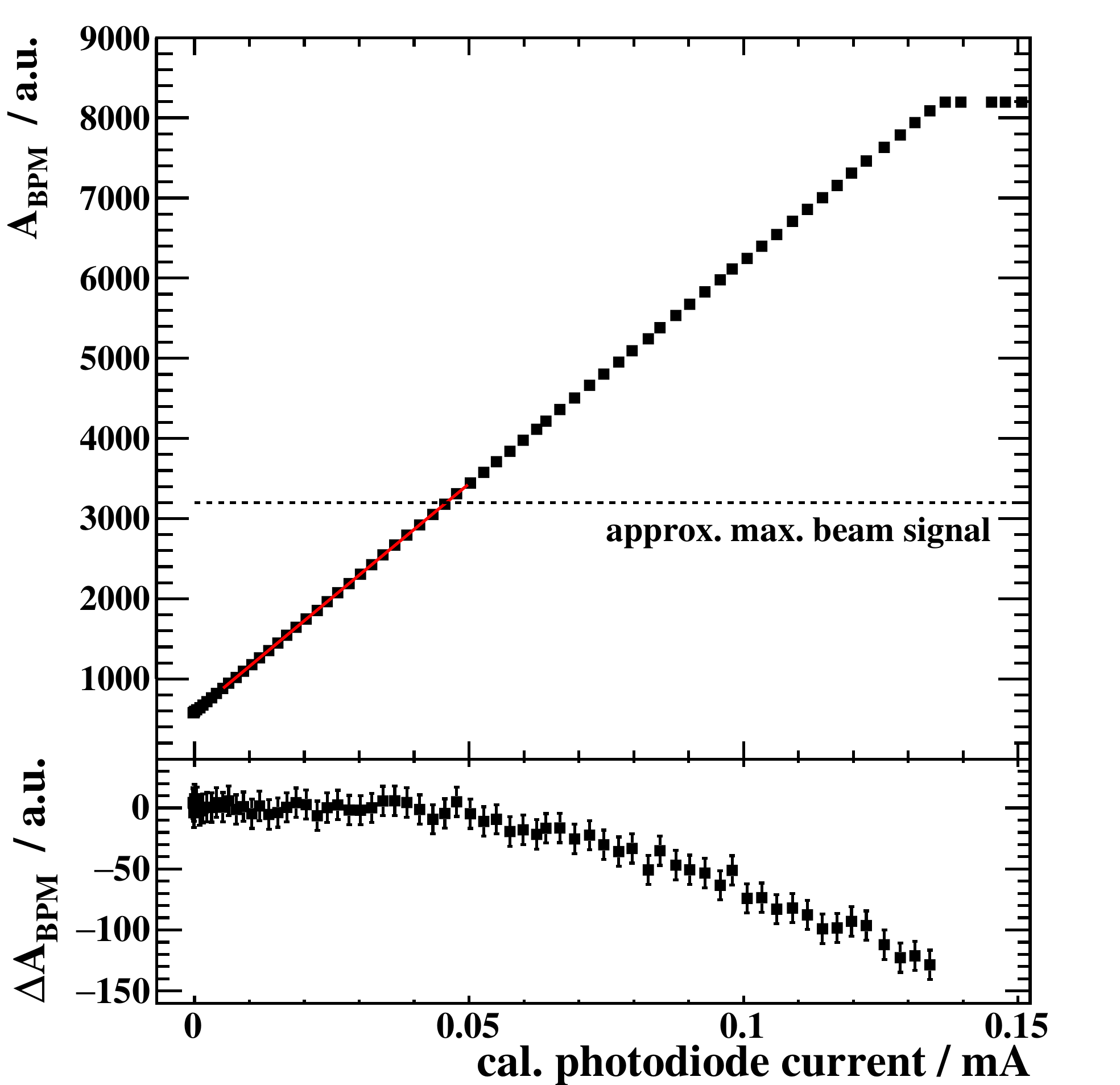}
  \caption{The linearity of the scintillating fibres coupled to the silicon arrays is shown in the top plot for a \SI{370}{\nm} UV light-source. The light intensity is measured with a calibrated photodiode for a single channel.  The maximum signal amplitude seen in beam data, resulting from a high fluctuation in a proton spill with insensity setting I9 (\SI{2E9}{\per\s}),  is overlaid for comparison. A linear approximation is based on the lower third of the data (solid line). The difference of the data with respect to the approximated line is shown in the bottom plot.  }
  \label{fig:uvlinearity}
\end{figure}

A UV(\SI{380}{\nm}) LED was used to excite the wavelength shifting component of the scintillator while a calibrated photodiode\footnote{Newport 818-UV} was used  to monitor the intensity of the LED. In Figure~\ref{fig:uvlinearity}, the relative measured intensity in ADC counts in the central channel is plotted versus the current measured in the calibrated photodiode. The response of the scintillating fibres and the photodiode arrays together is quite linear over the range of interest, but begins to deviate from linearity on the percent level towards the saturation limit of the ADC. This is likely due to anon-negligible internal resistance in series with the photodiodes. The maximum signal observed from the ion beam is indicated for comparison, showing that the system is a safe distance from saturation when the integration window is the nominal \SI{100}{\mics} in length while having a moderate dynamic range. The electrical noise observed in the system during this test produces a pedestal width\footnote{one sigma} of 12~ADC channels.

\subsection{Beamline Measurement Setup}
The experimental room at HIT provides a horizontal beam with a laser alignment system indicating the iso-center, where the beam properties are regularly checked during quality assurance measurements, and targets are typically placed. The BPM detector planes have also been placed at this location.

In some measurements two detector planes have been used. The second plane was placed a short distance behind the first which was at the iso-center. The additional readout was daisy-chained to the first and resulted in a reduction in overall readout rate due to the limited bandwidth of the available electronics.

\begin{figure}[htbp]
\centering
  \includegraphics[width=0.6\linewidth]{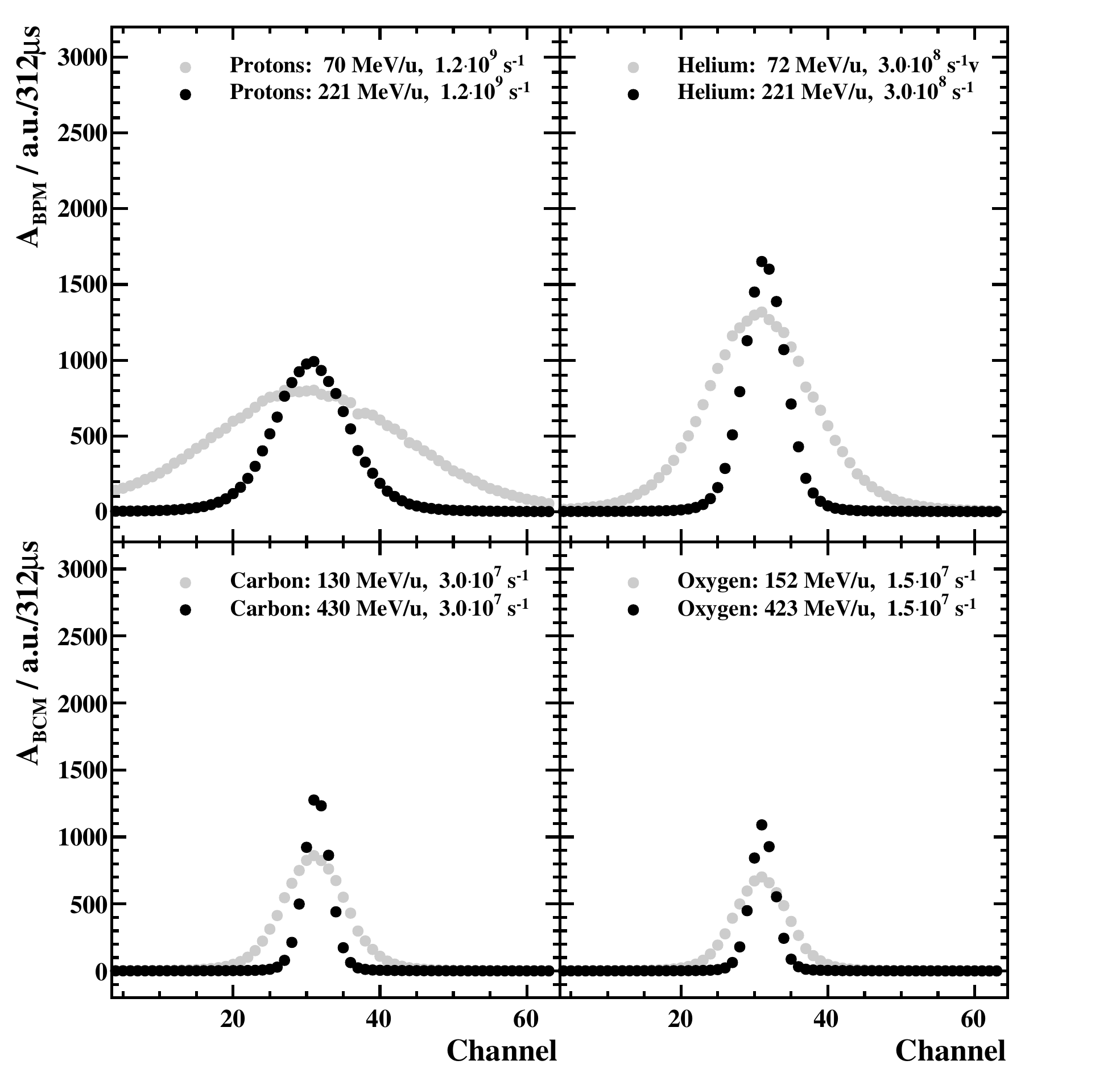}
  \caption{ The observed signal amplitudes averaged over two spills for \SI{312}{\micro\s} integration windows with proton, helium, carbon and oxygen beams for two energy settings (highest and lowest with stable intensity) at a single intensity setting, I8.  The absolute energy and intesity values are given in the plots.  }
  \label{fig:beamprofileexample}
\end{figure}

 The collected data has the form of beam profiles collected at a given frequency. The average beam profiles  for proton, helium, carbon and oxygen beams are shown in Figure~\ref{fig:beamprofileexample} for the highest available energy setting as well as the lowest setting where the intensity is stable. An intensity setting of I8 was used for all the examples, though the average fluence depends on the ion species and is given in the data plot along with the beam energy used.

\subsection{Response to Ion Beam Intensity}
\begin{figure}[htbp]
  \centering

  \mbox{}\hfill  

  \begin{subfigure}[t]{.49\linewidth}
	\centering\includegraphics[width=0.98\linewidth]{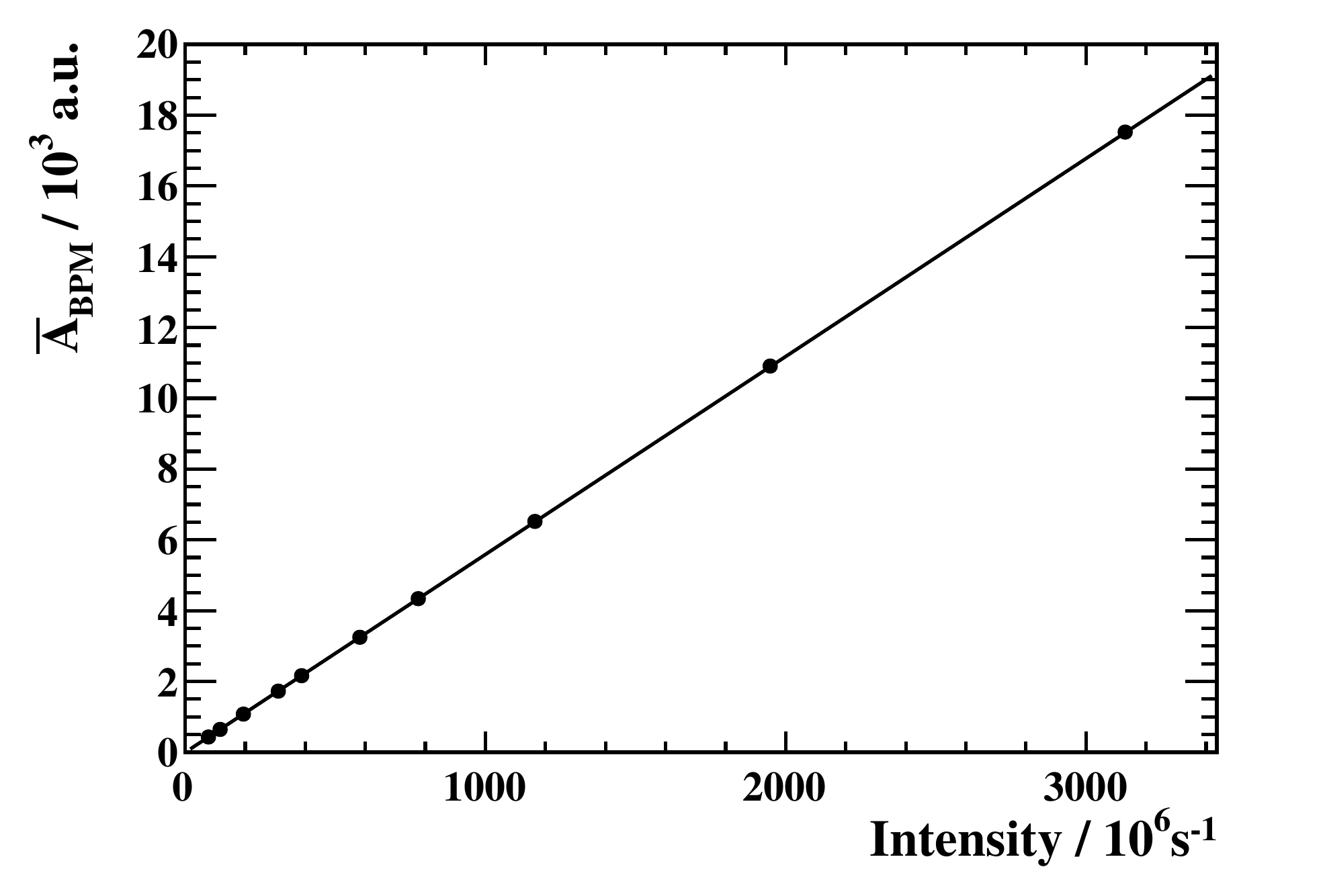}
    \caption{Protons}
    \label{fig:protonIntlinearity}
  \end{subfigure}\hfill
  \begin{subfigure}[t]{.49\linewidth}
     \centering\includegraphics[width=0.98\linewidth]{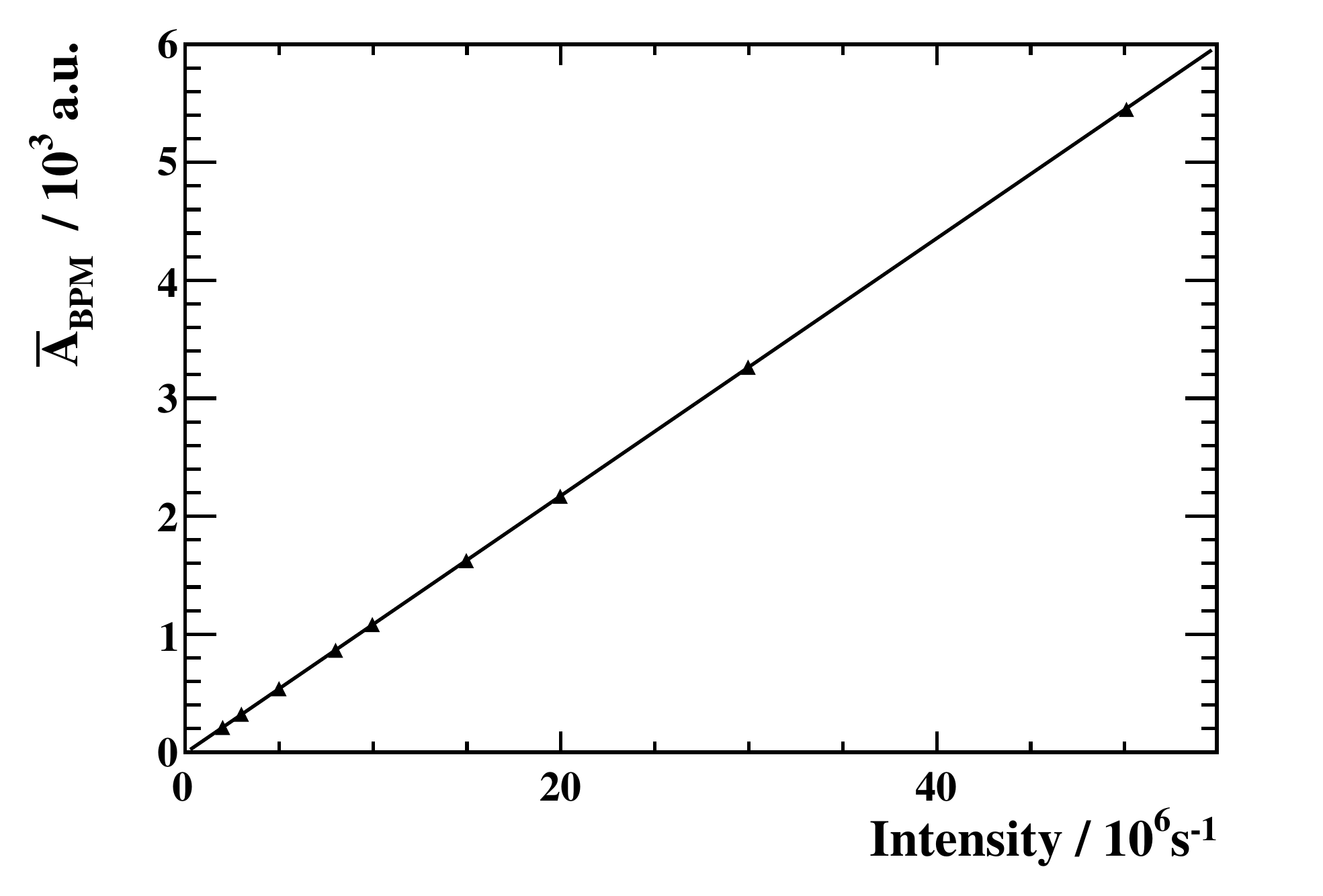}
    \caption{Carbon Ions}
    \label{fig:carbonIntlinearity}
  \end{subfigure}
  \hfill\mbox{}
\caption{  \label{fig:Intlinearity} The relative mean signal amplitude scanning over the available intensities at HIT for (a) \SI{221}{\mega\electronvolt}/u protons (\SI{8E7}-\SI{3.2E9}{\per\s}) and (b)  \SI{430}{\mega\electronvolt}/u carbon ions (\SI{2E6}-\SI{5E7}{\per\s}).}
\end{figure}

The intensity of the beam can be controlled by the user at the HIT clinic with up to ten available settings currently. For protons and carbon, this corresponds respectively to  \SI{8E7}-\SI{3.2E9}{\per\s} (I1-I0) and  \SI{2E6}-\SI{5E7}{\per\s} (I1-I9). The relative signal amplitude for the selected range of proton and carbon ion fluences is shown in Figures~\ref{fig:protonIntlinearity} and \ref{fig:carbonIntlinearity}. An integration time of \SI{312}{\mics} has been used with a low gain setting. A straight line is fit to the data to demonstrate the excellent linearity of the detector. This result is not entirely unexpected as the maximum density of particles is on the order of  \SI{1}{\ns^{-1}\cm^{-2}} with a scintillator decay time of \SI{2.36}{\ns} \cite{bib:borshchev}. The molecular excitation of the scintillator is highly localised and the probability of the scintillator being in an excited state is nearly zero for the next incident particle.

\subsection{Response to Ion Beam Energy and Species}

It is well known that the light yield of plastic scintillator is dependent upon the dE/dx of the particle depositing energy in the base scintillator. However, the literature available for this particular scintillating fibre does not extend to heavier ions. The measurements here will show the relative signal response of the \SI{0.250}{\mm} diameter Kuraray SCSF-78MJ scintillating fibres to the available ions species  (protons, helium, carbon, and oxygen ions) over their available range of energies. The stopping powers have been calculated using the PSTAR and ASTAR databases provided by NIST for protons and helium in polystyrene\cite{bib:astar, bib:pstar}. Stopping powers for carbon and oxygen are derived from the MSTAR database\cite{bib:mstar}.  

\begin{figure}[htbp]
  \centering

  \mbox{}\hfill  

  \begin{subfigure}[t]{.49\linewidth}
	\centering\includegraphics[width=0.98\linewidth]{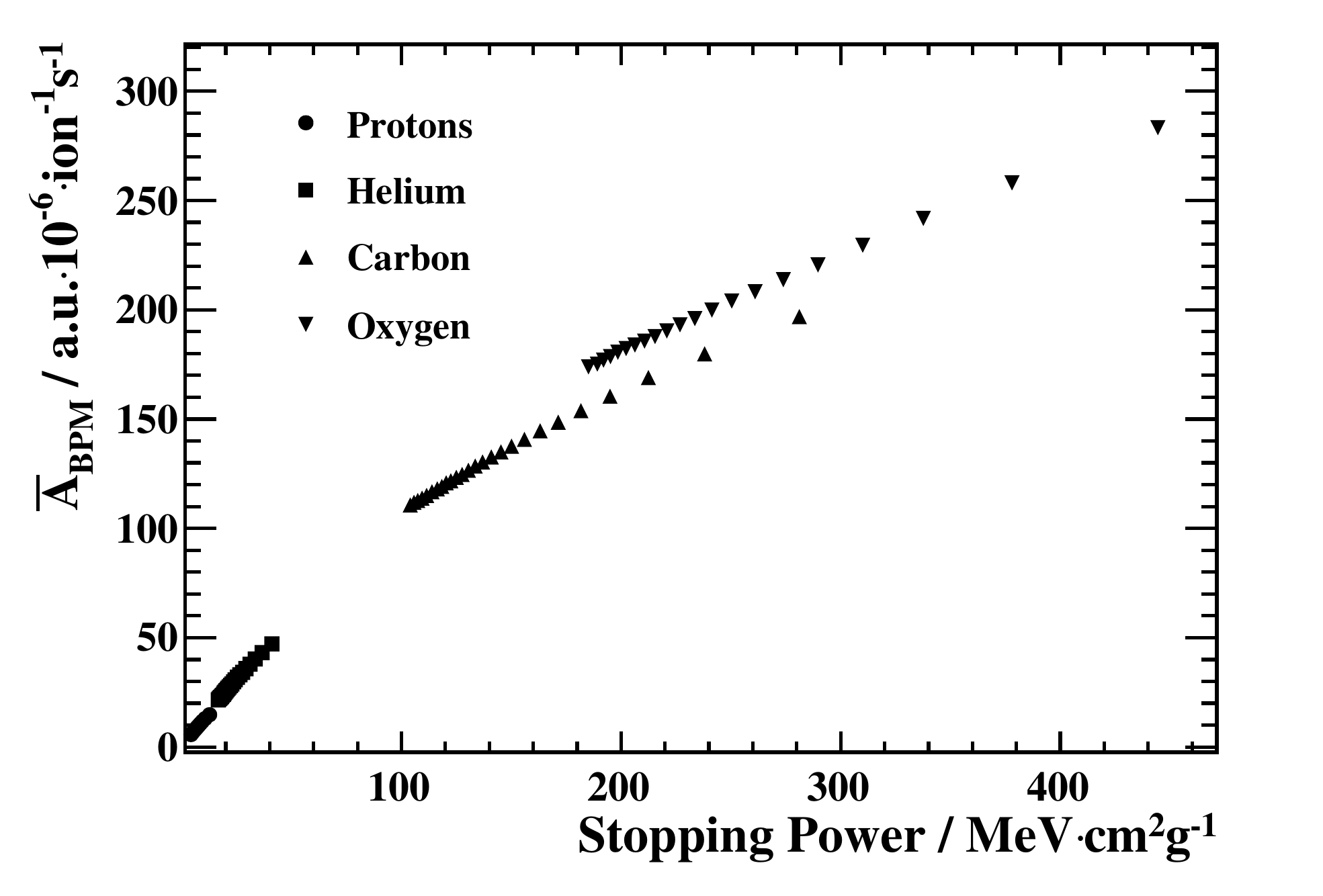}
    \caption{Linear Scale}
    \label{fig:signalVSsp-a}
  \end{subfigure}\hfill
  \begin{subfigure}[t]{.49\linewidth}
     \centering\includegraphics[width=0.98\linewidth]{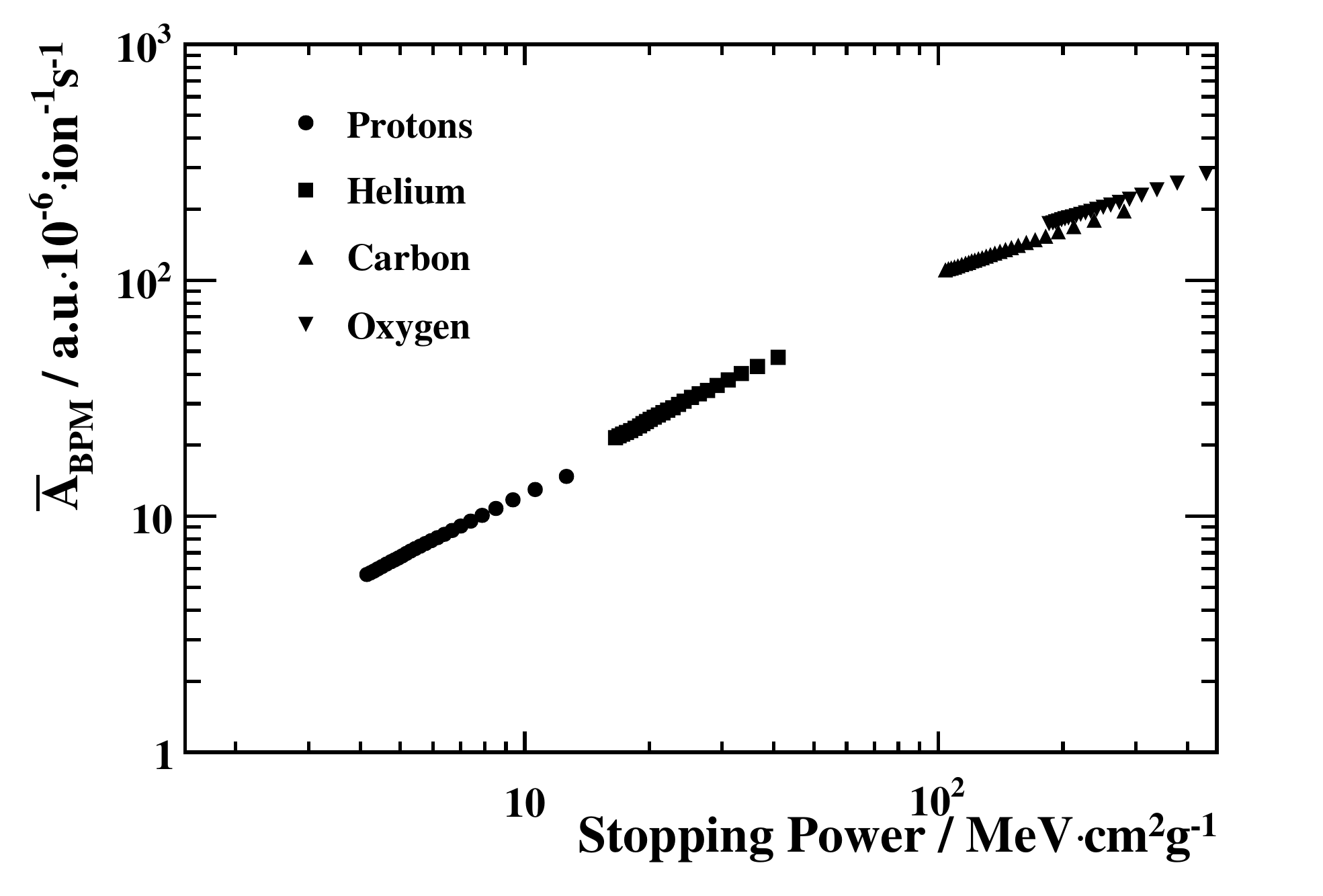}
    \caption{Log-Log Scale}
    \label{fig:signalVSsp-b}
  \end{subfigure}
  \hfill\mbox{}
\caption{  \label{fig:signalVSsp} The signal amplitude per particle per second for the four different ion types plotted as a function of their stopping power in polystyrene. }
\end{figure}

The measured light yield response of the scintillating fibre is plotted as a function of the stopping power in polystyrene in Figure~\ref{fig:signalVSsp-a} on a linear scale. The same is visible in  Figure~\ref{fig:signalVSsp-b} on a log-log scale to make the proton and helium data more visible.

Several models exist that describe the non-linear features observed in ion-induced scintillation. Birks first proposed that the specific scintillation signal amplitude per ion, $A_{ion}$, as a function of the stopping power can be described by \cite{bib:birks}:

\begin{equation}
\frac{A_{ion}}{dx} = C_{B}\frac{dE/dx}{1+k_{B}\cdot(dE/dx)},
\label{eq:birks}
\end{equation}

where $k_{B}$ is generally referred to as the quenching parameter, and is known to be ion charge dependent. $C_{B}$ is a signal gain parameter. A combined least-squares fit to the data in Figure~\ref{fig:signalVSsp-a}  with Eq.\ref{eq:birks} using the Minuit2 package of ROOT6 \cite{bib:root} results in the values listed in Table~\ref{tab:fitpars}. The values for  $k_{B}$ are consistent with others found in the other literature cited here, between  2 and \SI[inter-unit-product =$\cdot$]{20}{\mg\cdot\mega\electronvolt^{-1}\cm^{-2}} and increases for decreasing $Z$. 

Additional models developed since Birks' first measurements describe the energy deposition and saturation as the result of physical processes, typically described by a core region, in which the scintillator becomes saturated, and a halo, in which higher energy delta electrons have escaped the saturated core and the light yield is again linear to the deposited energy. One such model, introduced by Tarl\'{e} \cite{bib:tarle} and Ahlen \cite{bib:ahlen}, following the work of Voltz et al. \cite{bib:voltz}, describes the scintillation yield as:

\begin{equation}
\frac{A_{ion}}{dx} = C_{BVT}\frac{dE}{dx}\cdot \left(\frac{1-F_{h}}{1+B_{S}(1-F_{h})dE/dx} + F_{h}\right)  ,
\label{eq:BVT}
\end{equation}

where $C_{BVT}$ is a gain parameter and $B_{S}$ is the linear saturation coefficient. This will be referred to later as the BVT model. The fraction of scintillation light produced in the halo, $F_{h}$, is a function of the Lorentz parameters $\beta$ and $\gamma$, as shown in Eq.~\ref{eq:fh}, with an adjusted ionisation potential, $I_{ps}=68.7$~eV, for polystyrene, and a free parameter, $T_{0}$, which can be interpreted as something like the minimum energy of the delta rays needed to escape the saturated core.

\begin{equation}
F_{h} = \frac{1}{2}\frac{ ln (2mc^{2}\beta^{2}\gamma^{2}/T_{0}) - \beta^{2}}{ ln (2mc^{2}\beta^{2}\gamma^{2}/I_{ps}) - \beta^{2} }
\label{eq:fh}
\end{equation}

\begin{table}[htbp]
\centering

\begin{tabular}{@{}l|l|l@{}}
\toprule
        & Birks Model                                 & BVT Model                          \\ \midrule
    Protons & $k_{B} = 17.8\pm0.5~ mg\cdot MeV^{-1}cm^{-2}$  & $T_{0} = 1.3\pm0.4~GeV$                    \\
Helium  & $k_{B} = 6.8\pm0.2~ mg\cdot MeV^{-1}cm^{-2}$   & $T_{0} = 280\pm56~keV$                     \\
Carbon  & $k_{B}= 3.89\pm0.05~ mg\cdot MeV^{-1}cm^{-2}$  & $T_{0} = 33\pm4~keV$                       \\
Oxygen  & $k_{B} = 3.11\pm0.02~ mg\cdot MeV^{-1}cm^{-2}$ & $T_{0} = 15\pm2~keV$                       \\
        &                                              & $B_{S} = 5.4\pm0.2~mg\cdot~MeV^{-1}cm^{-2}$ \\
        &            $\chi^{2}/ndf = 191/94 $                                 &                   $\chi^{2}/ndf = 142/93  $     \\  \bottomrule
\end{tabular}
\caption{\label{tab:fitpars} The results of fitting Eq.~\ref{eq:birks} and Eq.~\ref{eq:BVT} to the data in Figure~\ref{fig:bvtbeta}. Each function is simultaneously fit to the four ion data sets. Similar gain constants of $C_{B}=(1.461\pm0.005)$ and $C_{BVT}=(1.425\pm0.004)  \cdot \mathrm{10^{-6} a.u./(MeV\cdot ion \cdot s^{-1}}$) are found in the fits.  }
\end{table}

\begin{figure}[htbp]
\centering
  \includegraphics[width=0.75\linewidth]{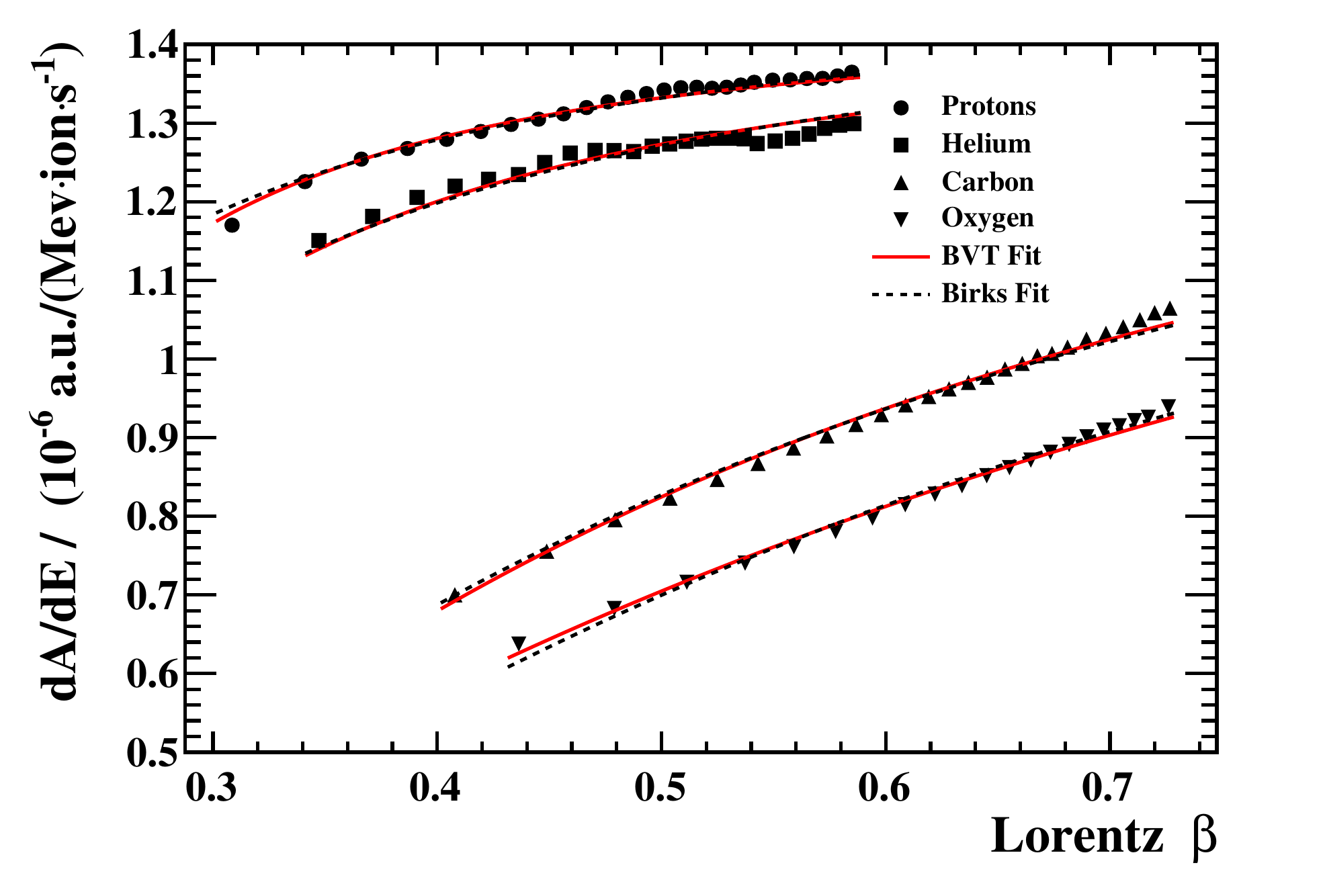}
  \caption{The scintillation yield per $10^{6}$ ions per second per unit energy deposited as a function of Lorentz $\beta$ for protons, helium, carbon, and oxygen ions measured at HIT.}
  \label{fig:bvtbeta}
\end{figure}

Given that both $F_{h}$ and the stopping power, $dE/dx$, are dependent only on the ion charge and Lorentz $\beta$ (Bethe-Bloch), it makes more sense to examine $dA/dE = dA/dx \cdot dx/dE$ as a function of $\beta$, as shown in Figure~\ref{fig:bvtbeta}. A combined fit of all four data sets to Eq.~\ref{eq:BVT} produces a single value of $B_{S} = 5.4\pm0.2~\mathrm{mg\cdot MeV^{-1}cm^{-2}}$. This is consistent with the result from Broggio et al.\cite{bib:broggio} where fixed energy helium (150 MeV/u), carbon (290 MeV/u) and neon (400 MeV/u) ions were incident on a vinyltoluene plastic scintillator, and PMMA blocks were used to reduce the ion energy before the scintillator. The combined fit  to our data was also constrained to a single value of $C_{BVT}$ and produces a reasonable  $\chi^{2}/ndf = 142/93  $, which is slightly smaller than that of the Birks Model fit.

 Requiring a single value of $T_{0}$ for all data sets did not produce a good result when fit to our data, especially the proton data.  As such, $T_{0}$  was left as a free parameter for each data set. It can be seen that the value for oxygen agrees well with the other large $Z$ measurements in \cite{bib:ahlen} and \cite{bib:broggio}. However, the value for protons appears unphysical, as the kinematic limit for delta ray production by heavy charged particles is $T_{max} = 2m_{e}c^{2}\beta^{2}\gamma^{2} < 600$~keV for protons with $\beta<0.6$. Despite this,  the increasing values for $T_{0}$ for decreasing $Z$ is consistent with the core-halo model, such that increasing values of $T_{0}$ indicate that an increasing amount of the scintillation light is produced in the saturated core region and increasingly higher energy delta electrons are required to reach an unsaturated halo region.  In the case of protons, this would mean that the entire region is saturated as no delta rays are energetic enough to escape to an unsaturated halo and the saturation per unit energy deposited is much larger than for higher $Z$ ions, as can be observed in the matching Birks coefficient, $k_{B}$.

\subsection{Detector Performance}
 In the following measurements, two detector planes were used with a readout rate of \SI{1}{\kilo\Hz} and an integration time close to the desired operational value of \SI{100}{\mics} with the high gain setting. Data was only collected for proton and carbon ion beams in this measurement as they are the two ion species used for patient treatment so far. The detectors are placed with the first plane at the isocenter with the second \SI{2}{\cm} behind.

\subsubsection{Signal-to-noise Ratio}
\begin{figure}[htbp]
  \centering

  \mbox{}\hfill  

  \begin{subfigure}[t]{.49\linewidth}
	\centering\includegraphics[width=0.98\linewidth]{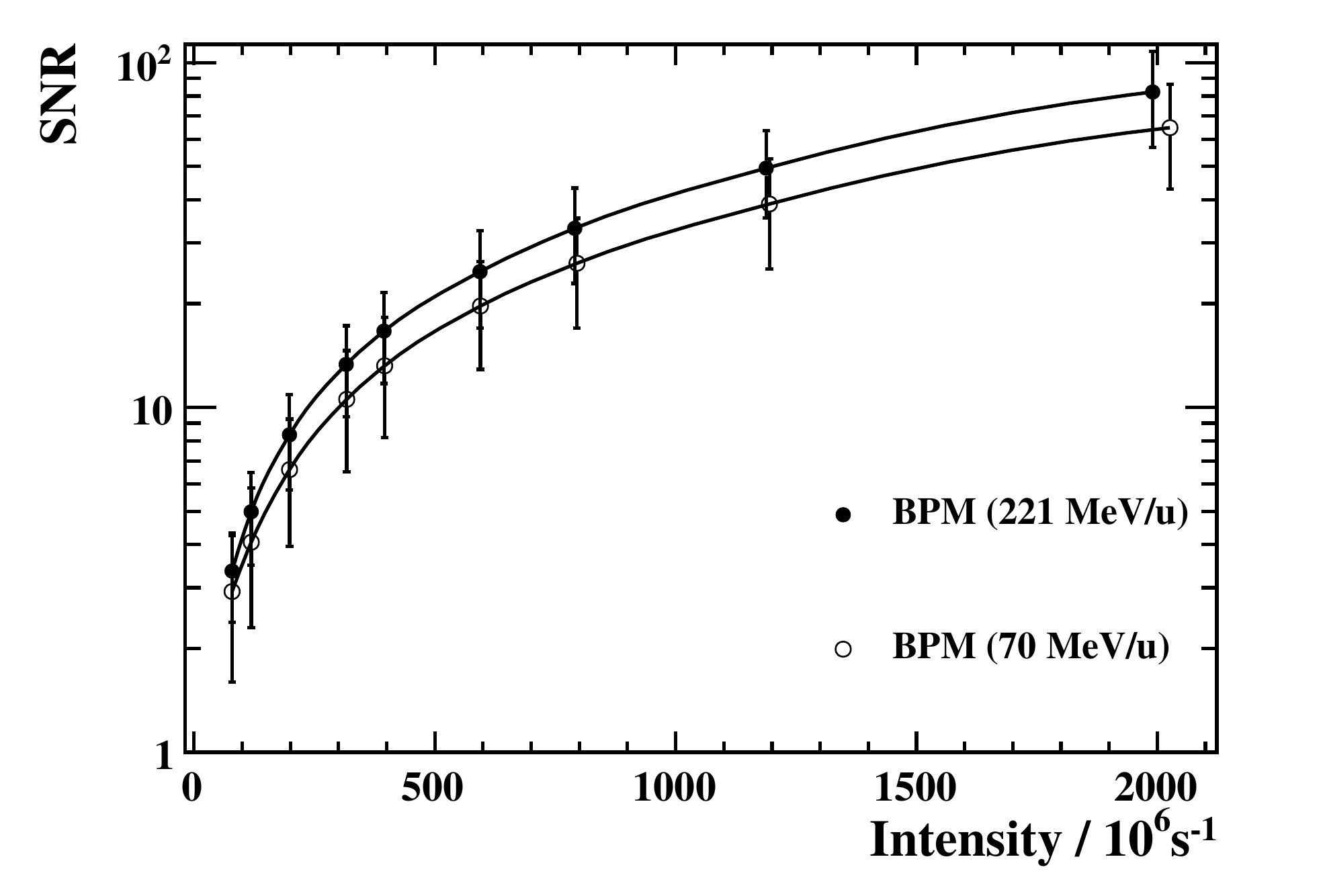}
    \caption{Protons}
    \label{fig:SNR-protons}
  \end{subfigure}\hfill
  \begin{subfigure}[t]{.49\linewidth}
     \centering\includegraphics[width=0.98\linewidth]{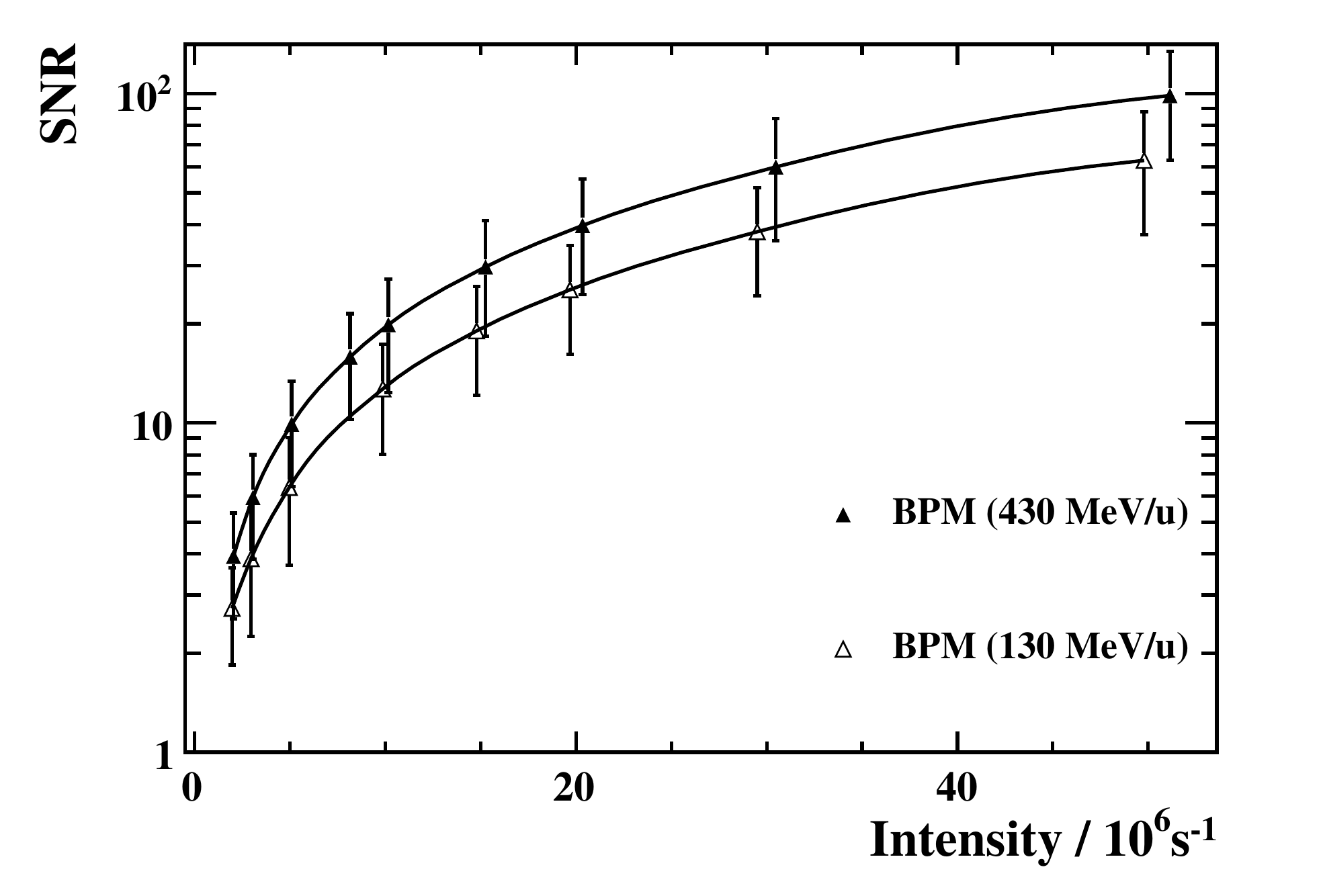}
    \caption{Carbon ions}
    \label{fig:SNR-carbon}
  \end{subfigure}
  \hfill\mbox{}
\caption{  \label{fig:SNR} The signal-to-noise ratio over the range of available intensities  for (a) 70 and 221 MeV/u protons and (b) 130 and 430 MeV/u carbon ions at the isocenter. An integration period  of \SI{100}{\mics} was used with a high gain setting. The error bars are the RMS of the peak amplitude distribution to $\sigma_{noise}$ ratio, which is dominated by the beam structure over time. The wider beam at the iso-center for the lower energy ions produces a worse SNR for the same intensity despite the slightly larger dE/dx. }
\end{figure}

The signal-to-noise ratio (SNR) of the detector is an important parameter in being able to extract correctly the beam position, intensity and width. Here we have defined the SNR as half the mean peak amplitude for each intensity setting divided by the standard deviation of the baseline noise, $\sigma_{noise}$. The position downstream of the currently installed tracking detectors will result in a wider beam  (due to multiple scattering) relative to  the few mm beam width at the MWPC locations. This naturally decreases the SNR such that an improved value would be expected in real use. In Figure~\ref{fig:SNR}, the lowest three intensity settings have an SNR less than five and the performance will likely be adversely affected by the electronics noise for the beam settings measured. The error bars in the figure are the RMS of the peak amplitude distribution to  $\sigma_{noise}$ ratio, such that, even for higher intensities, some integration windows will have an SNR less than five. \\

Improved dedicated electronics and additional shielding will likely improve the SNR in a further iteration of this detector system. A preliminary design has indicated a factor of five improvement is possible.  Moving the detector further upstream would also result in more light per channel  along with the use of a mirror.  An increased integration period could also improve the SNR and would be feasible at low intensities but would limit the readout rate of the beam position to a value similar to the MWPCs. \\

\subsubsection{Beam Position Resolution}
\begin{figure}[htbp]
  \centering

  \mbox{}\hfill  

  \begin{subfigure}[t]{.49\linewidth}
	\centering\includegraphics[width=0.98\linewidth]{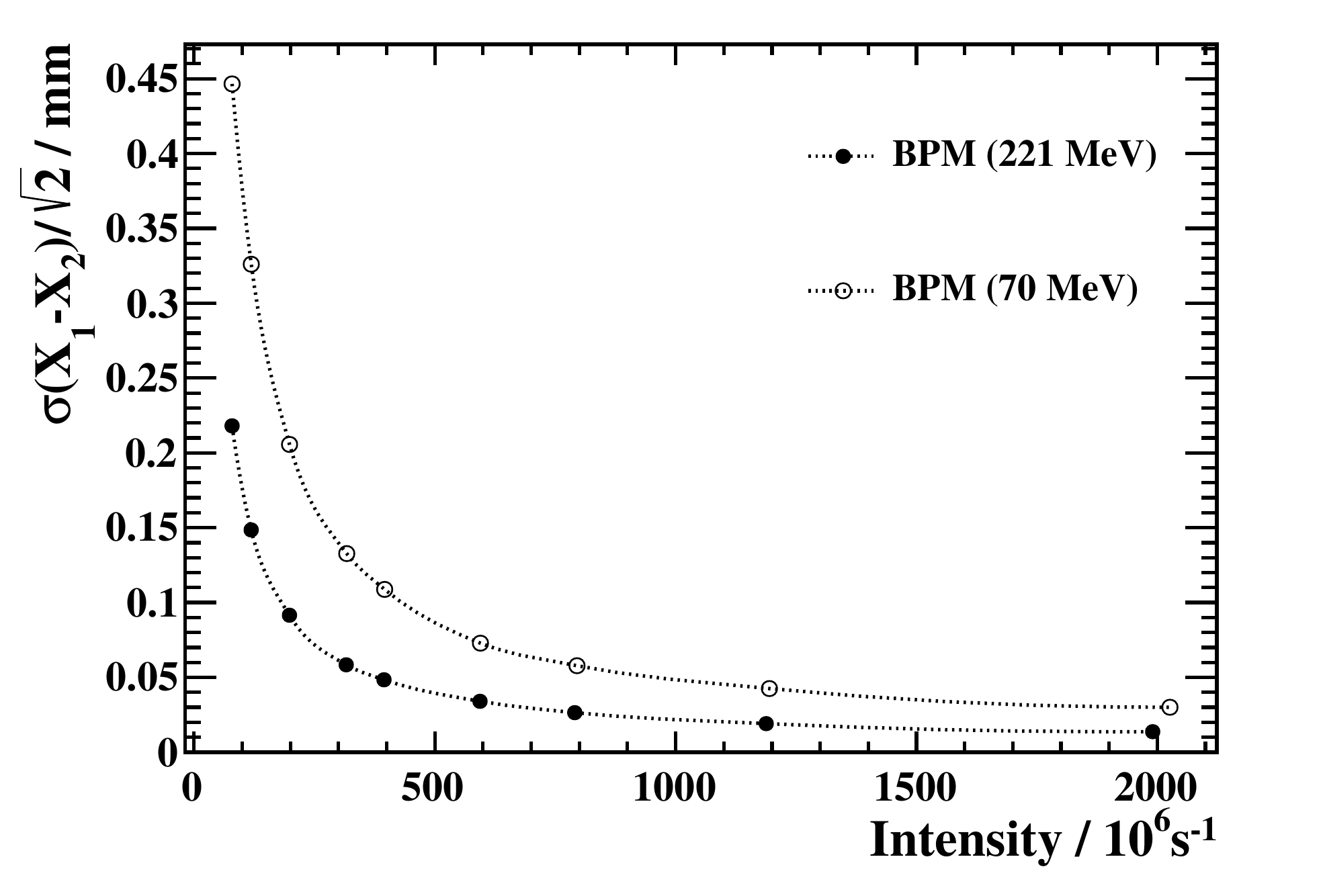}
    \caption{Protons}
    \label{fig:posdiff-protons}
  \end{subfigure}\hfill
  \begin{subfigure}[t]{.49\linewidth}
     \centering\includegraphics[width=0.98\linewidth]{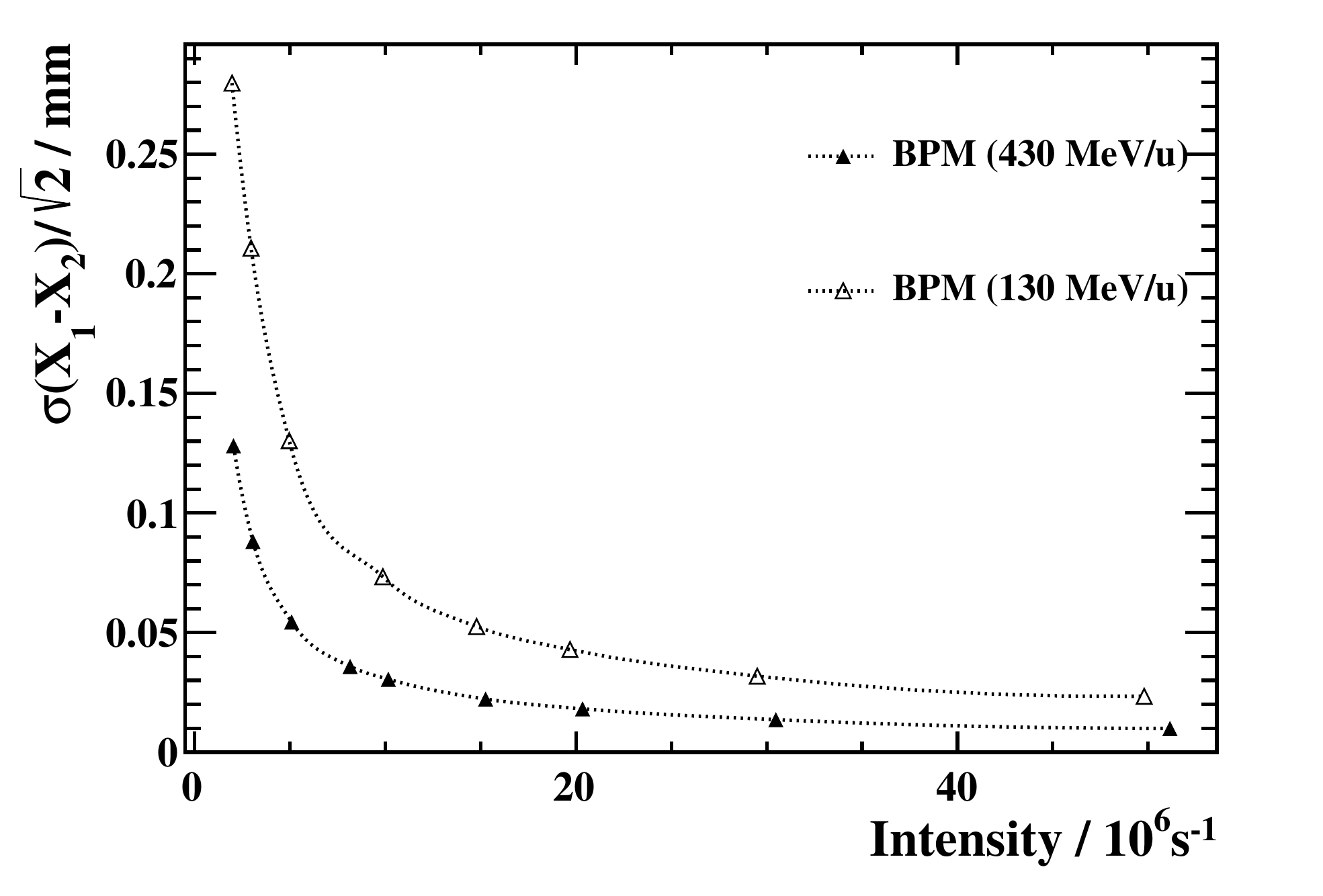}
    \caption{Carbon ions}
    \label{fig:posdiff-carbon}
  \end{subfigure}
  \hfill\mbox{}
\caption{  \label{fig:posdiff} The beam position resolution over the range of available intensities  for (a) 70 and 221 MeV/u protons and (b) 130 and 430 MeV/u carbon ions at the isocenter.   An integration period  of \SI{100}{\mics} was used with a high gain setting.}
\end{figure}

The beam position in each detector plane is determined from a Gaussian fit to the detector data and compared to the second detector plane. 
The per plane beam position resolution is given in Figure~\ref{fig:posdiff} for proton and carbon ion beams where the detector has been placed at the treatment iso-center.  The clinic requires that any new tracking system be able to resolve a beam with a FWHM of about \SI{1}{mm} with a position resolution better than \SI{0.2}{\milli\metre}.  The distance further downstream from the nominal tracking station positions means the beam has broadened significantly and achieving this with a low energy proton beam is not likely. The broad beam and the low SNR results in a position resolution of \SI{0.45}{\milli\metre} for the lowest intensity and low energy proton beam, improving to less than  \SI{30}{\micro\metre} for the highest intensity. The highest energy has a top resolution of \SI{12}{\micro\metre}.  
 The beam profile for carbon ions is less affected by multiple scattering in the detector systems, and, as such, is a better representation of the performance of the BPM as a tracking detector. A position resolution of  \SI{0.28}{\milli\metre} is achieved for the lowest intensity and low energy carbon ion beam and the highest intensity and energy achieves a position resolution of \SI{10}{\micro\metre}. Future measurements will attempt to move as far as possible upstream to achieve a more realistic estimate of the final working conditions.

\subsubsection{Beam Width Resolution}
\begin{figure}[htbp]
  \centering

  \mbox{}\hfill  

  \begin{subfigure}[t]{.49\linewidth}
	\centering\includegraphics[width=0.98\linewidth]{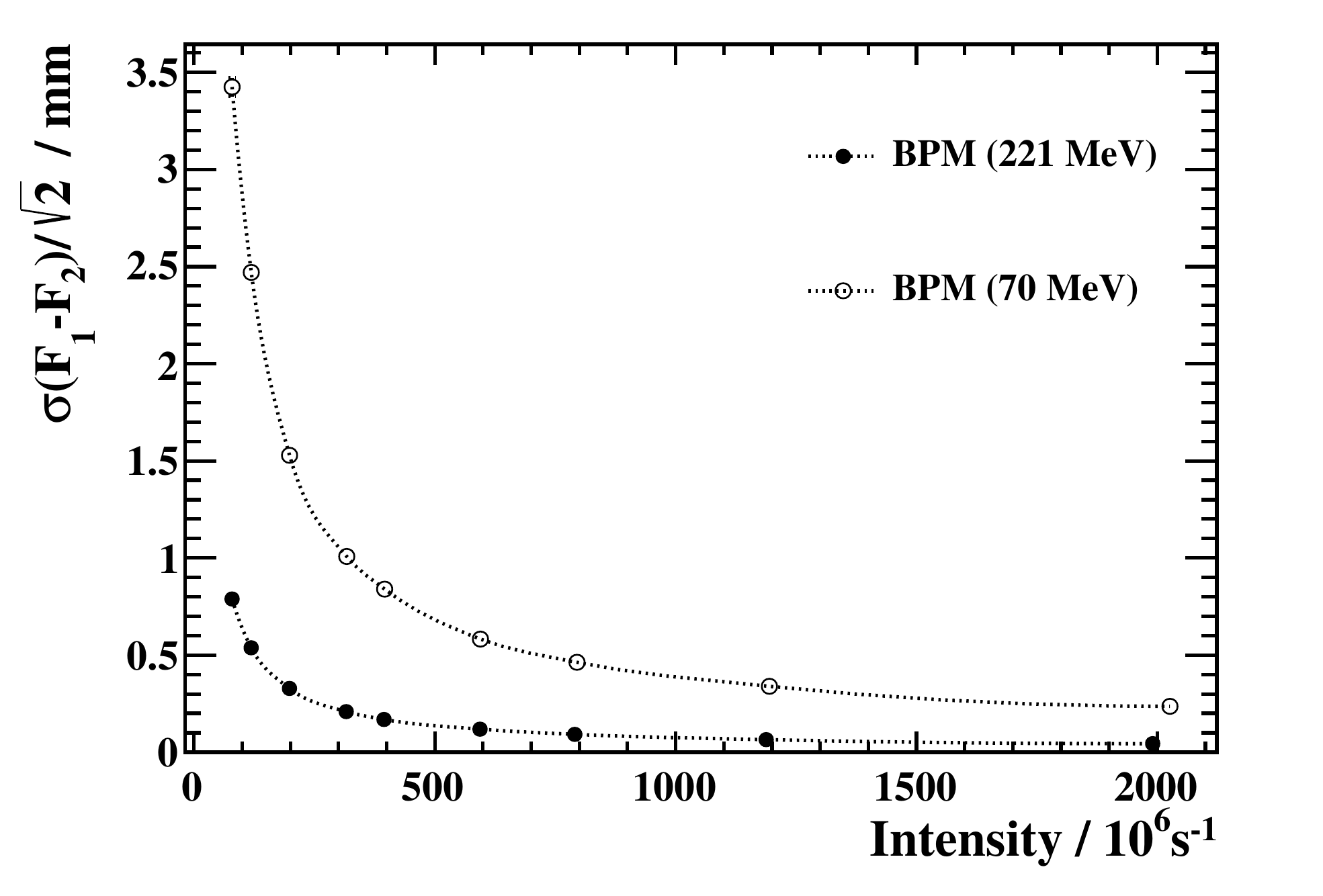}
    \caption{Protons}
    \label{fig:focusdiff-protons}
  \end{subfigure}\hfill
  \begin{subfigure}[t]{.49\linewidth}
     \centering\includegraphics[width=0.98\linewidth]{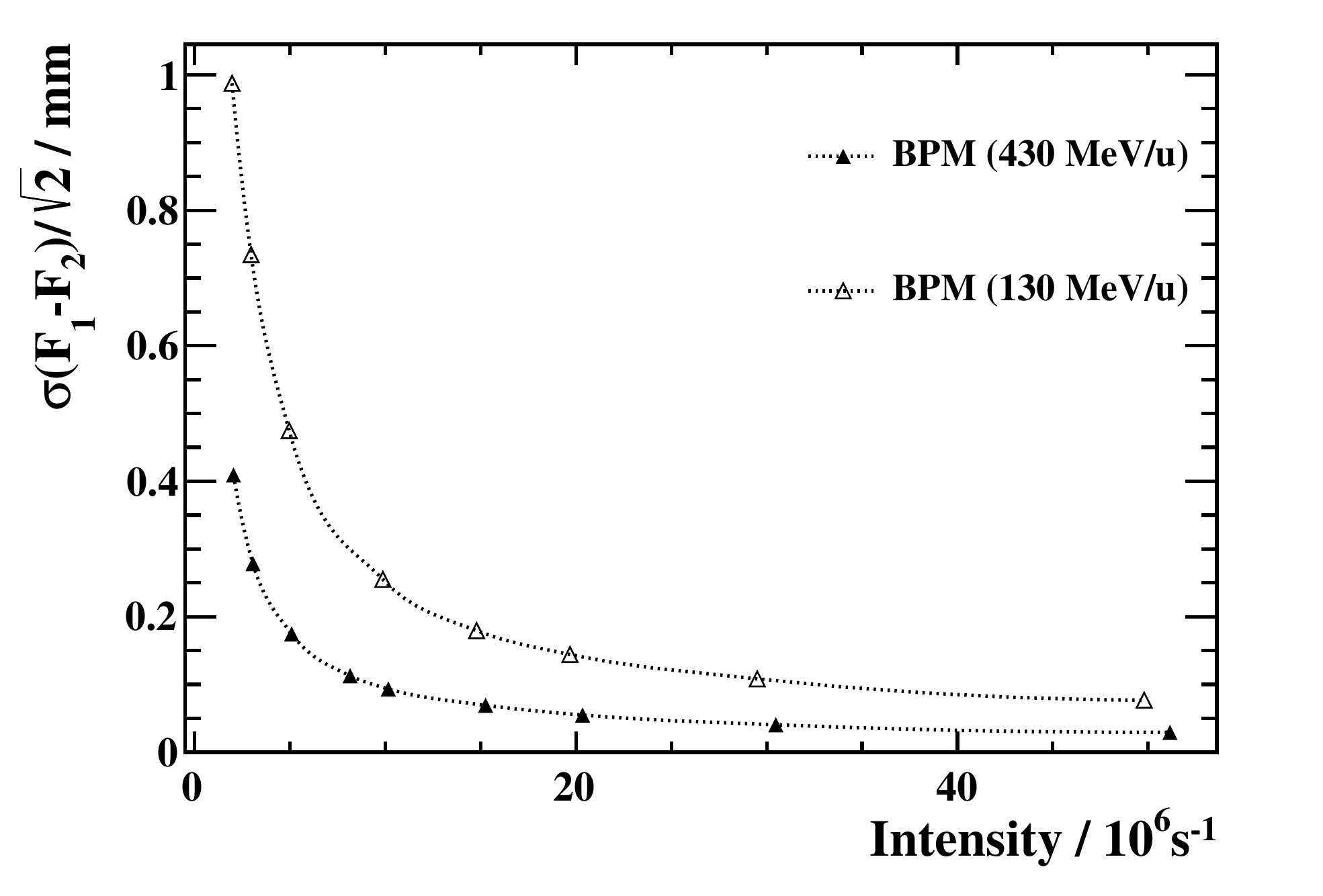}
    \caption{Carbon ions}
    \label{fig:focusdiff-carbon}
  \end{subfigure}
  \hfill\mbox{}
\caption{  \label{fig:focusdiff} The absolute beam width/focus (FWHM) resolution over the range of available intensities  for (a) 70 and 221 MeV/u protons and (b) 130 and 430 MeV/u carbon ions at the isocenter.  An integration period  of \SI{100}{\mics} was used with a high gain setting. }
\end{figure}

Correctly reconstructing the beam width is as important for patient treatment as the beam position, as the correct distribution of dose amongst voxels needs to be determined. The clinic has asked that the width resolution be determined better than \SI{0.4}{\milli\metre} for the tracking stations. The standard deviation of the difference in the measured beam width between the two detector planes, $\sigma(F_{1}-F_{2})/\sqrt{2}$   is shown in Figure~\ref{fig:focusdiff}. As can be seen in the figure,  $\sigma(F_{1}-F_{2})/\sqrt{2}=3.5$~mm for a single detector plane for the lowest intensity \SI{70}{\mega\electronvolt} protons and  the highest intensity \SI{221}{\mega\electronvolt} proton beam width has a resolution of \SI{0.044}{\mm}. In general, the low intensity proton beam width is not reconstructed as well as desired. This is again dominated by the poor SNR and the beam being broader than the detector. The narrower carbon beams are better reconstructed with a resolution of   $\sigma(F_{1}-F_{2})/\sqrt{2}=$~\SI{0.98}{\mm} for low intensity \SI{130}{\mega\electronvolt}/u  ions  and, for high intensity \SI{430}{\mega\electronvolt}/u ions,  the width resolution is as low as $\sigma(F_{1}-F_{2})/\sqrt{2}=0.03$~mm. Much like the beam position resolution, these values will likely improve as the tracking planes are moved further upstream, the absolute width of the beam narrows, and the SNR improves.

\subsubsection{Signal Amplitude Resolution}

\begin{figure}[htbp]
  \centering

  \mbox{}\hfill  

  \begin{subfigure}[t]{.49\linewidth}
	\centering\includegraphics[width=0.98\linewidth]{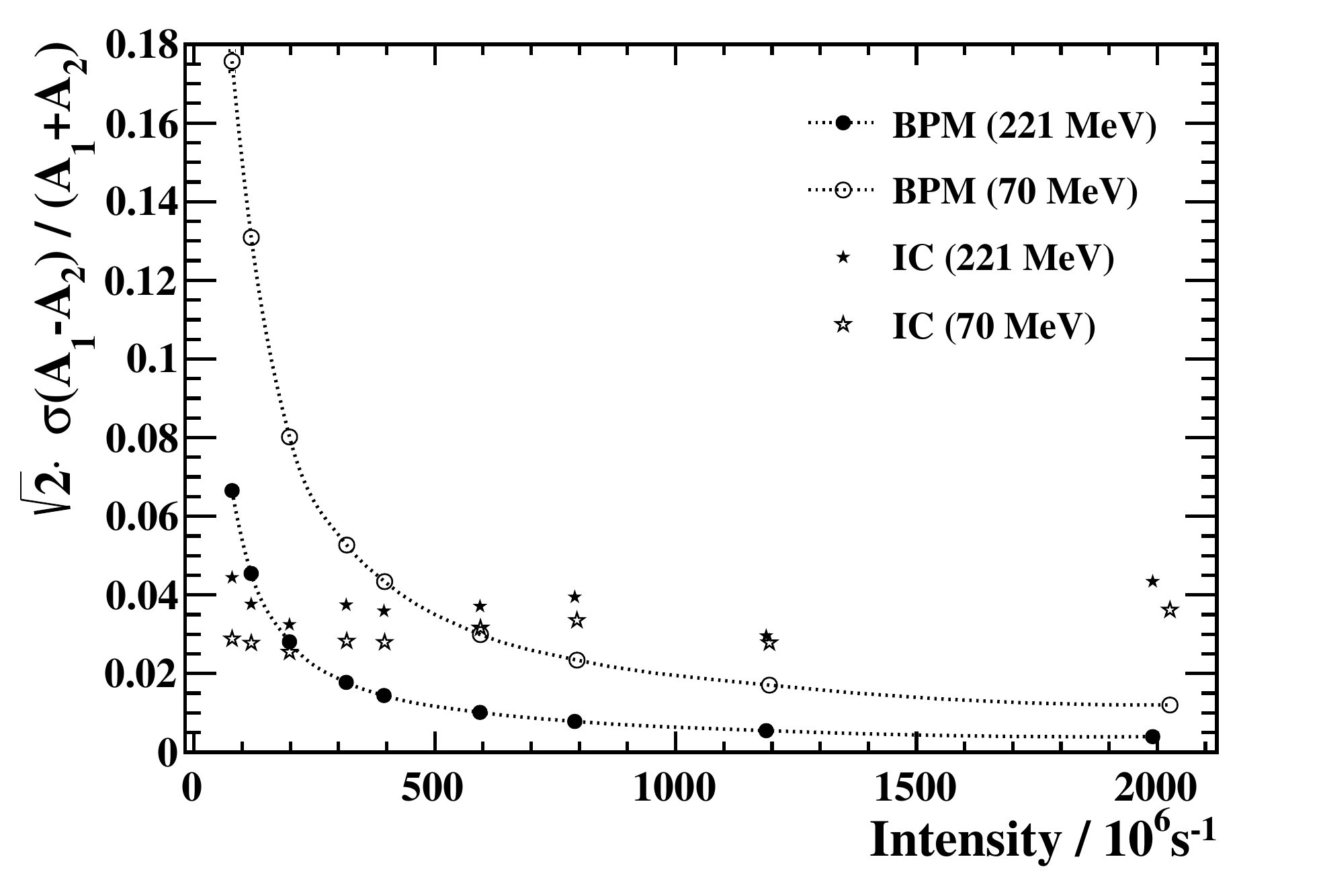}
    \caption{Protons}
    \label{fig:sigdiff-protons}
  \end{subfigure}\hfill
  \begin{subfigure}[t]{.49\linewidth}
     \centering\includegraphics[width=0.98\linewidth]{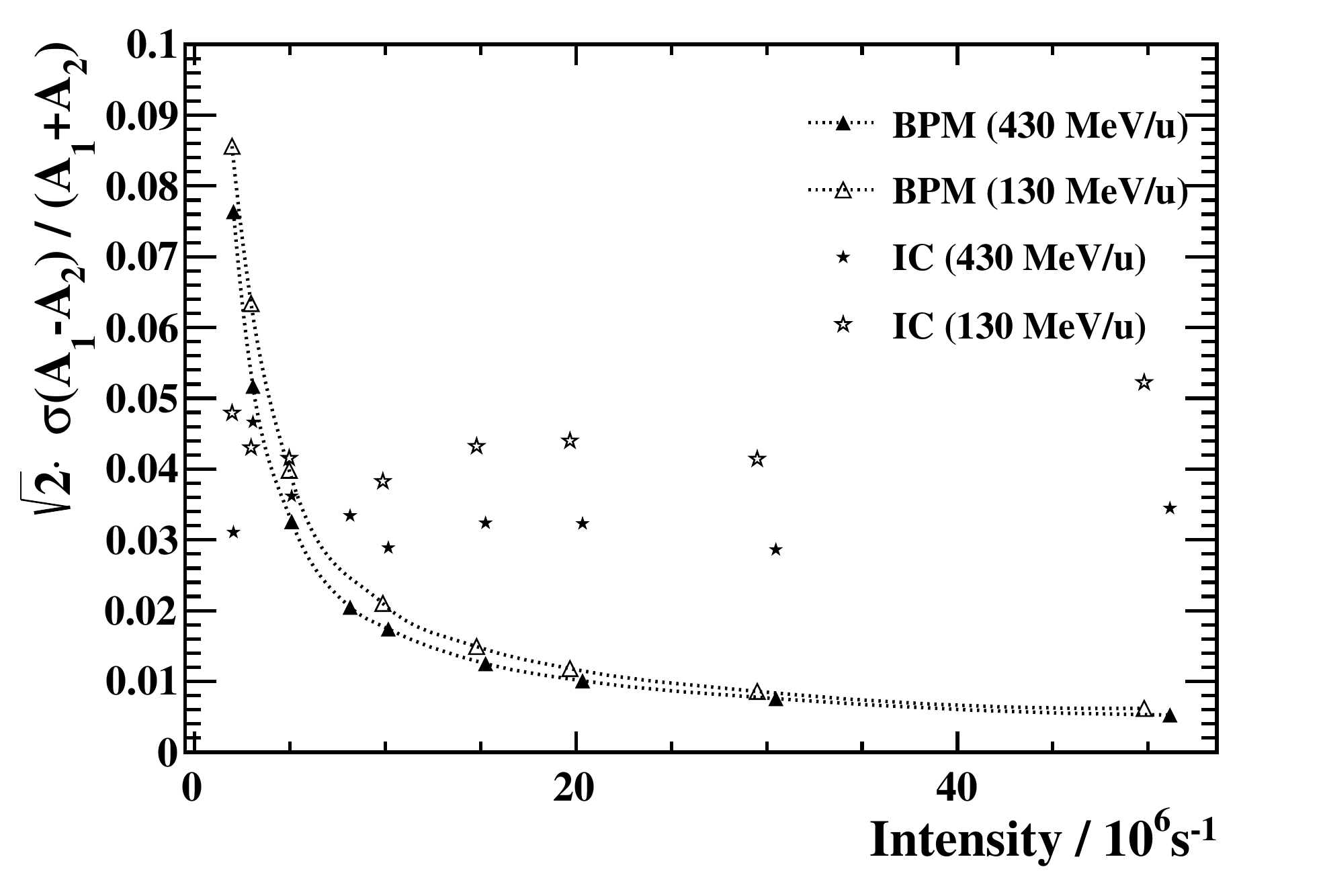}
    \caption{Carbon ions}
    \label{fig:sigdiff-carbon}
  \end{subfigure}
  \hfill\mbox{}
\caption{  \label{fig:sigdiff} The signal resolution over the range of available intensities  for (a) 70 and 221 MeV/u protons and (b) 130 and 430 MeV/u carbon ions at the isocenter.   An integration period  of \SI{100}{\mics} was used with a high gain setting.}
\end{figure}

The purpose of the BPM is to primarily monitor the position and width of the beam. However, given the predictable response of the scintillator to each ion type and intensity, it may be possible to use the detector as a secondary dose monitor complimentary to the ionisation chambers already installed. A measure of the dose measurement performance is the signal resolution. 
The signal resolution in each \SI{100}{\mics} integration window, or the sigma of the difference in signal amplitude\footnote{After correction for a small difference in gain.} divided by the mean signal amplitude plotted as a function of the beam intensity, is shown in Figure~\ref{fig:sigdiff} for protons and carbon ions. A small correction is applied to the gain of the second detector plane amplitude to provide the same mean signal amplitude. The additional division by $\sqrt{2}$ provides the single detector plane resolution, as the two planes are identical in design and construction with some residual differences remaining in the fibre light yield, transmission, and the photodiode array gain. For comparison, the same calculation for the ionisation chamber (IC) data provided by the EtherCAT system at HIT is plotted. Here, the ionisation current from two ionisation chambers is provided in \SI{50}{\mics} intervals via EtherCat. A timestamp is injected into the asynchronously operated readout systems (BPM and EtherCAT) for offline time alignment. The IC data is then averaged over the equivalent photodiode integration window for this analysis. For operational and safety purposes at the Clinic, the IC current is monitored via a different system with faster time scale on the order of \SI{1}{\mics}\footnote{It is potentially feasible to implement a fast current readout on the photodiode arrays as well.}. 

As can been seen in Figures~\ref{fig:sigdiff-protons} and \ref{fig:sigdiff-carbon}, the performance is strongly dependent on the intensity of the beam. As expected from the SNR, the lower intensities suffer in performance but for the majority of the intensity settings, the signal resolution is comparable to or better than that observed in the IC data. However, the IC data is quite flat over the range of intensities. Ideally, the total dose for each treatment voxel  should be known to the order of 0.5\% and with the treatment time per voxel on the order of milliseconds, the resolution per \SI{100}{\mics} integration should be much better than 10\% and closer to 2--3\%. For high intensities, the current state of the BPM is sufficient. However, scientists from the clinic have communicated that the intensity used during patient treatment is typically at the lower end of the scale. Improvements in the SNR from increased integration times, as well as reduced electronics noise, will directly improve the signal resolution in future developments. 

\section{Towards a light two-layer fibre ribbon}
\label{section:geant}

\begin{figure}[htbp]
\centering
  \includegraphics[width=0.75\linewidth]{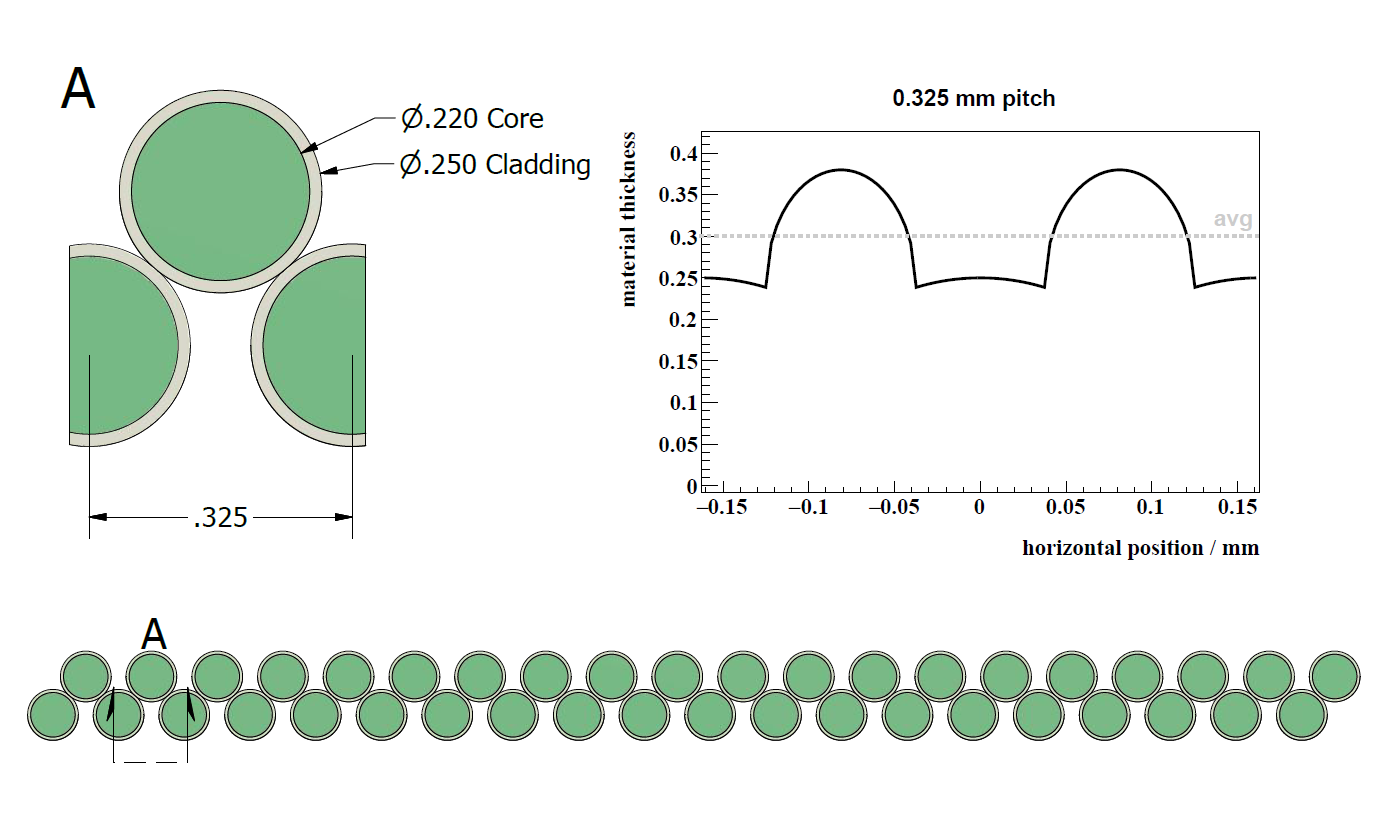}
  \caption{A proposed two layer mat configuration without glue. The total material thickness across one unit of pitch (\SI{0.325}{\mm}) for a two layer mat is shown in the plot (solid black line). The average material thickness is indicated with the dashed gray line. }
  \label{fig:twolayer}
\end{figure}

As stated earlier, the minimum achievable thickness for a two-layer fibre mat without glue is on average \SI{0.3}{\mm}. This is slightly more than the fibre diameter due to the overlap of the two fibre layers, as seen in Figure~\ref{fig:twolayer}. The thickness of the material that would be traversed by an ionised particle varies between \SI{0.24}{\mm} and \SI{0.38}{mm} with a frequency of one half of the fibre pitch. This pitch in the figure shown is \SI{0.325}{\mm}, which is slightly more than the mats currently produced for LHCb. A new winding technique would likely need to be developed in order to construct the ribbon without glue in the active area.

\subsection{A simple GEANT4 study}

A very simple Monte Carlo simulation study was performed to develop a preliminary understanding of the effect on the pristine beam when additional material such as proposed fibre tracker is introduced into the HIT beamline. Following the example of Parodi \textit{et al} \cite{Parodi_2012}, a beamline geometry description was used consisting of equivalent water layers of varying thickness representing the material of each of the three ionisation chambers (\SI{0.230}{\mm}) and each of the two MWPCs (\SI{0.130}{\mm}), along with the vacuum window (\SI{0.320}{\mm}). It was also suggested to use \SI{3}{\micro\m} of tungsten to correctly estimate the lateral scattering of low momentum protons due to the wires of the MWPC.  The vacuum window is placed at $z=\SI{-1280}{\mm}$ with  the isocenter at $z=0$. The layers for the BAMS stations are placed at  $z= (-1260, -1210, -1160, -1110, -1060)$~mm, an approximation of the current system based on Ref.~\cite{bib:ringbaek}. No ripple filter is modelled, though this is typically used to increase the spread of the Bragg peak position during treatment with carbon ions.  The simulated beam begins in vacuum before the vacuum window with a FWHM of \SI{2.5}{\mm}, corresponding to a focus setting of F1 at HIT. No beam divergence is applied.  The lateral spread of 10k ions was then scored at the isocenter.  For comparison with a scintillating fibre based tracking system, the MWPC water and tungsten layers are replaced with a \SI{0.6}{\mm} water layer. The recommended physics libraries in GEANT4~\cite{Agostinelli:2002hh} for the \textit{/advanced/hadrontherapy} example are used including electromagnetic and hadronic scattering routines.  

 In general, the simulation of the HIT beamline with MWPC layers underestimates the lateral spread by 20\% compared to the measured values and those listed in the HIT beam library, and is worse at lower energies.  However, to establish the impact of additional material in the beamline, a relative comparison is made between the current beamline material (MWPC equiv. layers) and the increased material of the scintillating fibre planes. The lateral spread of the beam for protons and carbon ions for four energies is shown in Table~\ref{tab:geant}. The relative increase in lateral spread at the isocenter ranges from 1.6\% for 400 MeV/u carbon ions up to 8.4\% for 50 MeV/u protons. Replacing the fibre station material with polystyrene results in a slightly smaller increase in lateral spread than with water, but the conclusions are the same even if a factor of two is assumed to account for any uncertainties in the simulation.  The impact on the spread of the beam is low, less than 10\% in this simulation,  and the pull-back of the Bragg peak could be easily accounted for. 

\begin{table}[]
\centering
\caption{The lateral spread of the ion beam (FWHM) scored at the isocenter in a GEANT4 based simulation of the HIT beamline for the current material description or with scintillating fibre based tracking. }
\label{tab:geant}
\begin{tabular}{@{}llll@{}}
\toprule
Ion Type            & w/MWPC  (mm) &w/SciFi (mm) & Ratio \\ \midrule
Protons (50 MeV/u)  & 25.01             & 27.11             & 1.084 \\
Protons (100 MeV/u) & 13.12             & 13.97             & 1.065 \\
Protons (150 MeV/u) & 9.09              & 9.63              & 1.060 \\
Protons (200 MeV/u) & 6.95              & 7.49              & 1.078 \\ \midrule
Carbon (100 MeV/u)  & 6.05              & 6.55              & 1.082 \\
Carbon (200 MeV/u)  & 3.78              & 4.00              & 1.060 \\
Carbon (300 MeV/u)  & 3.18              & 3.34              & 1.052 \\
Carbon (400 MeV/u)  & 2.99              & 3.04              & 1.016 \\ \bottomrule
\end{tabular}
\end{table}

\section{Radiation Damage}

An additional challenge to this detector is the loss of transmission due to radiation damage in the plastic scintillating fibres. While the total dose expected in one year over the surface of the fibre area is relatively low from patient treatment (a few kGy), the total fluence through the central point of the detector plane from regular accelerator activities is very high, >\SI{1}{\mega\gray}, producing a darkened spot a couple mm in diameter.    It is well known that the transmission of plastic scintillator is reduced with a strong wavelength dependence and has been measured up to \SI{35}{\kilo\Gray} for this particular fibre \cite{bib:scintrad}. Primarily the blue component is blocked leading to the yellowing of the fibres. As the decay times of the all fibres are very fast compared to the integration window we intend to use, a slower but  more radiation hard fibre like 3HF will be investigated. The 3HF fibres emit light with a peak wavelength of \SI{530}{\nm} which also matches better with the spectral sensitivity of the photodiode which peaks at \SI{700}{\nm}. Additionally, an optical filter may reduce the sensitivity to the change in certain wavelengths due to radiation damage. A mirror would also increase the total light yield collected at the photodiode further improving the SNR.

\section{Calibration}

A detector of this style will likely need to be calibrated for two reasons. The first is due to the natural attenuation length of the optical signal in the fibres. These fibres have a long attenuation length component of over \SI{3}{\m} with a shorter component on the order of \SI{30}{cm}. The addition of a mirror largely removes the dependence on position away from the readout end with the signal differing only by up to 10-20\% across the length of the fibre. The second reason is the possibility of damage or natural variation amongst fibres. Certain channels might produce a smaller optical signal thereby distorting the measured light distribution shape over the channels. The calculated barycentre as well as RMS would then be biased to a certain degree depending on the relative position of the beam over the damaged channel. The plan will be to use a fine grid of uniform beam spots over the detector to determine the calibration factors at regular intervals. The damaged region due to radiation damage would also be visible in the calibration study and the impact could be removed. This calibration procedure will be investigated in future studies.

\section{Conclusion}

We have measured the scintillation response of Kuraray SCSF-78MJ scintillating fibres to proton, helium, carbon and oxygen ion beams over the range of energies and intensities used at the Heidelberg Ion Therapy Clinic. The linearity and dynamic range of the fibre and photodiodes have been shown to be suitable for the measurements at the desired integration times. Measurements of the beam properties and the residual distributions also show that, while the detector does not currently surpass the requirements at low intensities, it is a feasible set of technologies that could be developed further to meet the specifications of the HIT clinic.  An improved SNR will result from the addition of a mirror, as well as sampling the narrower beam further upstream. Dedicated electronics rather than an adapted commercial product should significantly reduce the electronics noise. Producing ribbons with only two layers of scintillating fibre will reduce the material in the beam to acceptable levels.

\acknowledgments

We wish to thank the staff and scientists of HIT for the fruitful and productive discussions regarding the detector development and for providing us with the beam time in the experimental room. We also wish to thank the LHCb Scintillating Fibre Tracker collaboration for providing us with a sample of a scintillating fibre ribbon for our measurements.



\end{document}